\documentclass{aa} 

\usepackage{txfonts}
\usepackage{longtable}  
\usepackage{natbib}
\usepackage{graphicx}
\usepackage{epstopdf}
\usepackage{amssymb}
\usepackage{hyperref}
\bibpunct{(}{)}{;}{a}{}{,}
\def\Teff{\ensuremath{T_{\mathrm{eff}}}}
\def\logTeff{\ensuremath{\log T_{\mathrm{eff}}}}
\def\LogTeff{\ensuremath{\textrm{Log}\,T_{\mathrm{eff}}}}
\def\logg{\ensuremath{\log g}}

\def\vsini{\ensuremath{{\upsilon}\sin i}}
\def\kms{$\mathrm{km\,s}^{-1}$}

\def\logl{\ensuremath{\log L/L_{\odot}}}

\begin{document}

\title{Pulsational properties of ten new slowly pulsating B stars}

\author{M.~Fedurco\inst{1}
\and E.~Paunzen\inst{2} 
\and S.~H{\"u}mmerich\inst{3,4}
\and K.~Bernhard\inst{3,4}
\and \v{S}.~Parimucha\inst{1}}
\institute{Institute of Physics, Faculty of Science, P. J. {\v S}af{\'a}rik University, Park Angelinum 9, Ko{\v s}ice 040\,01, Slovak Republic
\email{miroslav.fedurco@student.upjs.sk}
\and{Department of Theoretical Physics and Astrophysics, Masaryk University,
Kotl\'a\v{r}sk\'a 2, CZ-611\,37 Brno, Czech Republic}
\and{American Association of Variable Star Observers (AAVSO), 49 Bay State Rd, Cambridge, MA, 02138, USA}
\and{Bundesdeutsche Arbeitsgemeinschaft für Ver{\"a}nderliche Sterne e.V. (BAV), D-12169, Berlin, Germany}
}

\date{} 
\abstract
{Slowly pulsating B (SPB) stars are upper main-sequence multi-periodic pulsators that show non-radial g-mode oscillations driven by the $\kappa$ mechanism acting on the iron bump. These multi-periodic pulsators have great asteroseismic potential and can be employed for the calibration of stellar structure and evolution models of massive stars.}
{We collected a sample of ten hitherto unidentified SPB stars with the aim of describing their pulsational properties and identifying pulsational modes.}
{Photometric time series data from various surveys were collected and analyzed using diverse frequency search algorithms. We calculated astrophysical parameters and investigated the location of our sample stars in the \logTeff\ versus \logl\ diagram. Current pulsational models were calculated and used for the identification of pulsational modes in our sample stars. An extensive grid of stellar models along with their g-mode eigenfrequencies was calculated and subsequently cross-matched with the observed pulsational frequencies. The best-fit models were then used in an attempt to constrain stellar parameters such as mass, age, metallicity, and convective overshoot.}
{We present detected frequencies, corresponding g-mode identifications, and the masses and ages of the stellar models producing the best frequency cross-matches. We partially succeeded in constraining stellar parameters, in particular concerning mass and age. Where applicable, rotation periods have been derived from the spacing of triplet component frequencies. No evolved SPB stars are present in our sample. We identify two candidate high-metallicity objects (HD\,86424 and HD\,163285), one young SPB star (HD\,36999), and two candidate young SPB stars (HD\,61712 and HD\,61076).}
{We demonstrate the feasibility of using ground-based observations to perform basic asteroseismological analyses of SPB stars. Our results significantly enlarge the sample of known SPB stars with reliable pulsational mode identifications, which provides important input parameters for modeling attempts aiming to investigate the internal processes at work in upper main-sequence stars.}

\keywords{Asteroseismology -- stars: early-type -- stars: variables: general}

\titlerunning{ }
\authorrunning{Fedurco et al.}
\maketitle

\section{Introduction} \label{introduction}

Slowly pulsating B (SPB) stars were first described as a class by \citet{Waelkens91}. They are main-sequence (MS) stars of spectral types B2 to B9 (i.e., 22\,000 to 11\,000\,K), which show non-radial g-mode oscillations driven by the $\kappa$ mechanism acting on the iron bump \citep{gautschy93}. These SPB stars are multi-periodic pulsators whose observed periods range from 0.3 d to about 5\,d \citep{DeCat07}.

Several fast-rotating SPB stars have been described; the majority, however, seem to be comparably slow rotators \citep{DeCat07,Degroote11}. There also exist very slowly rotating SPB stars like KIC\,10526294 (rotation period of about 188\,d; \citealt{Papics14}). This is remarkable as the mean projected rotational velocity (\vsini) of B-type stars well exceeds 100\,\kms\ \citep{Abt02}, which induces strong meridional circulation and mass loss \citep{Dolginov83}.

Stars of the upper MS are characterized by a convective core and a radiative envelope. The asteroseismic potential of SPB stars has been recognized early on \citep{DeCat07}. These stars are perfect test laboratories of ill-understood processes that have a significant influences, such as diffusion and internal differential rotation, on the lifetime of a star. Asteroseismic analyses of SPB stars are therefore expected to contribute to the calibration of stellar structure and evolution models of massive stars, for which the observed mass distribution significantly contradicts theoretical predictions \citep{Castro14}. In addition, these analyses can be employed to determine and calibrate internal parameters such as the convective overshoot parameter or envelope mixing \citep{Moravveji2015, Moravveji2016}, which are not directly observable but have a significant influence on stellar structure and evolution \citep{Papics15, Buysschaert18}. Furthermore, nonrigid rotational profiles can be studied by determining the spacing between rotationally split modes \citep{Papics17, Triana15}.

Current asteroseismic modeling attempts are mainly based on quasi-uninterrupted observational data from space telescopes such as COROT (Convection, Rotation and planetary Transits)\citep{Corot09} or $Kepler$ \citep{Kepler16} because their continuity and precision enable reliable mode identification and help to constrain basic stellar parameters such as mass, luminosity, and effective temperature with great precision \citep{Szewczuk18}. However, owing to their long-period, gravity-driven oscillations, SPB stars lend themselves perfectly for asteroseismic modeling using ground-based survey data with rather low cadence.

Pulsation in pre-MS (PMS) stars is of special interest as it allows the investigation of the short-lived early phases of stellar evolution with oscillations \citep{Zwintz15}, thereby providing valuable constraints and input parameters for theoretical considerations. Dedicated effort has led to the discovery of $\gamma$ Doradus and, in particular, $\delta$ Scuti pulsation in PMS objects \citep[e.g.,][]{Zwintz13,Ripepi15}, but the situation is less clear for SPB stars. \citet{Gruber12} identify two SPB stars in the vicinity of the young open star cluster NGC\,2244; however, the available proper motion and radial velocity data were insufficient to confirm their cluster membership. \citet{Zwintz09} discovered ten SPB variables in the field of NGC\,2264, another young open cluster. Later on, \citet{Zwintz17} investigate four SPB variables belonging to NGC\,2264 in detail. Interestingly, despite the derived ages between one and six million years, the authors find that all stars seem to be early zero age MS (ZAMS) objects that have already left the PMS phase. The search for PMS SPB pulsators, therefore, has not yielded any conclusive results yet, and identifying suitable candidates remains of special interest.

In this paper, we present photometric time series analysis of ten newly identified SPB stars together with an asteroseismic analysis aimed at the identification of pulsational modes. These results significantly add to our knowledge of the pulsational properties of these stars and allow us to constrain astrophysical parameters, which helps to throw more light on the internal processes at work in upper MS stars.

\begin{table*}[t]
\caption{Astrophysical parameters of our sample stars.}
\label{tab:astro}
\begin{tabular}{lllllllllll}
\hline
HD & HIP/TYC & Spec. Type & \Teff & \logg & Parallax & $A_{\mathrm V}$ & $V$ & $M_{\mathrm V}$ & BC & \logl \\
& & & (K) & & (mas) & (mag) & (mag) & (mag) & (mag) \\
\hline
36999   & 4778-1364-1   & B7\,V & 13\,700       & 4.31 &        2.53(7) & 0.20(5) & 8.470(21) & +0.28(8) & $-$1.02 &      2.20(3) \\
48497   & 32221 & B5\,V & 14\,550       & 4.32 &        2.96(6) &       0.13(4) & 7.513(8)        & $-$0.26(6) & $-$1.17 &        2.47(2) \\
61076   & 36938 & B5/7\,III & 14\,900   & 4.42 &        2.15(3) &       0.39(8) & 9.130(17) &     +0.40(9) &      $-$1.23 &       2.23(4) \\
61712   & 37222 & B7/8\,V & 13\,900     & 4.28 &        2.00(4) & 0.23(10)      & 8.961(14) & +0.24(11) & $-$1.05 &       2.23(4) \\
66181   & 6558-2365-1   & B5\,V & 15\,450       & 4.35 &        1.93(12)        & 0.26(7) & 7.447(7) &    $-$1.39(15) &   $-$1.32 &       2.98(6) \\
86424   & 48811 & B9\,V & 12\,850       & 4.47 &        1.76(4) & 0.30(5)       & 9.298(14) &     +0.23(7) &      $-$0.86 & 2.15(3) \\
97895   & 54970 & B4\,V & 15\,050       & 4.39 &        1.44(7) & 0.32(16)      & 8.755(12) &     $-$0.77(19) &   $-$1.26 & 2.71(8) \\
115067  & 64658 & B8\,V & 13\,650       & 4.17 &        1.92(9) & 0.19(11)      & 8.062(12) &     $-$0.70(15) &   $-$1.01 & 2.59(6) \\
163285  & 87692 &  B8\,V & 13\,500      & 4.26 &        3.25(5) & 0.21(4)       & 7.737(8) &      +0.09(5) &      $-$0.98 & 2.26(2) \\
168121  & 89786 & B8/9\,III & 12\,600   & 4.26 &        2.62(8) &       0.41(12)        & 8.307(16) &     $-$0.01(13) &   $-$0.81 & 2.23(5) \\
\hline
\end{tabular}
\end{table*}

\section{Target selection, photometric data sources, reduction, and analysis} \label{observations}

The following sections give details on the sample selection, the employed photometric time series data, and our methods of analysis.

\subsection{Target selection and data sources} \label{sub_target_selection}
For the sample selection, we resorted to The International Variable Star Index (VSX; \citealt{Watson06}), which is the most up-to-date and accurate variable star database available. To identify new SPB stars, we systematically investigated suspected and known variable stars with periods typical for SPB stars and a spectral type of B or A. The extension to spectral type A was deemed necessary to identify objects that had been spectroscopically misclassified.

In the given spectral type range on the MS, SPB stars coexist with other types of photometric variables such as Be stars \citep{Rivinius13} and rotationally variable CP2/4 stars \citep{Preston74}. While Be stars usually exhibit complex variability on timescales ranging from a few minutes to decades, the latter objects, which are also known as $\alpha^{2}$ Canum Venaticorum (ACV) variables \citep{GCVS}, exhibit surface abundance patches or spots and variability periods in the same range as SPB stars. Their variability, however, is strictly monoperiodic \citep{Netopil17} and can therefore be easily distinguished from SPB type pulsation.

After the collection of an initial sample list, we searched for the availability of photometric observations in the databases of the following surveys: All Sky Automated Survey \citep[ASAS;][]{Pigulski14}, All-Sky Automated Survey for Supernovae \citep[ASAS-SN;][]{Kochanek17}, Hipparcos \citep{Leeuven97}, Optical Monitoring Camera \citep[OMC;][]{Alfonso15}, 
and Wide Angle Search for Planets \citep[SuperWASP;][]{Street03}. In this way, a sample of ten newly identified SPB variables boasting extensive time series photometry was collected (cf. Table \ref{tab:astro}).

HD\,48497, HD\,61076, HD\,61712, HD\,86424, HD\,115067, HD\,163285, and HD\,168121 were first reported as variables on the basis of Hipparcos data \citep{HIPPARCOS,Koen02}. \citet{Nichols10} report the variability of HD\,66181 using the pointing control camera of the Chandra X-ray Observatory. However, all these stars lack deeper studies and their variability types have not been determined. Consequently, the stars are listed as variable stars of unspecified type (type VAR) in the VSX. HD\,36999 and HD\,97895 are listed as suspected variable stars in the GCVS and VSX.

\subsection{Photometric data reduction and analysis} \label{sub_phot_data_reduction_and_analysis}

Only data from the third phase of the All Sky Automated Survey  project (ASAS-3) were taken into account; measurements with quality assignments ``C'' and ``D'' were excluded. Mean $V$ magnitudes were calculated as the weighted average of the values provided in the five different apertures available. To check the feasibility of this approach, we subsequently restricted our analysis to the ``best'' aperture as indicated by the ASAS-3 system for any given star. No significant differences were found between the two approaches. Finally, a basic 5$\sigma$ clipping was performed to clean the light curves from outliers.

The ASAS-SN measurements are taken with different cameras, which we treated separately. The mean for each individual data set was calculated, and data points were deleted on a 5$\sigma$ basis. After that, the data of the individual cameras were merged. In the case of Hipparcos and OMC data, a 5$\sigma$ clipping algorithm was applied.

Measurements with an error larger than 0.05\,mag were excluded from the SuperWASP data sets. Each camera was treated separately and the mean for each individual data set was derived; data points were deleted on a 5$\sigma$ basis. The data of the individual cameras were then merged. Applying this procedure also corrects for the different offsets of the cameras. However, for larger data sets, we also separately investigated the measurements from each camera.

The resulting light curves were examined in more detail using the program package PERIOD04 \citep{Lenz05}, which performs a discrete Fourier transform. An iterative pre-whitening procedure was used to extract all significant frequencies with a signal-to-noise ratio above 4. The results from PERIOD04 were checked with the CLEANEST and phase dispersion minimization (PDM) algorithms as implemented in the program package PERANSO \citep{Paunzen16}. The same results were obtained within the derived errors, which depend on the time series characteristics, i.e., the distribution of measurements over time and the photon noise. All significant frequencies are listed in Table \ref{tab:modes}. Amplitude spectra and residuals, as derived with PERIOD04, are given in the Appendix (Figure \ref{fig:amplitude_spectra}).

\begin{figure}[t]
 \centering
 \includegraphics[width=0.98\linewidth,keepaspectratio=true]{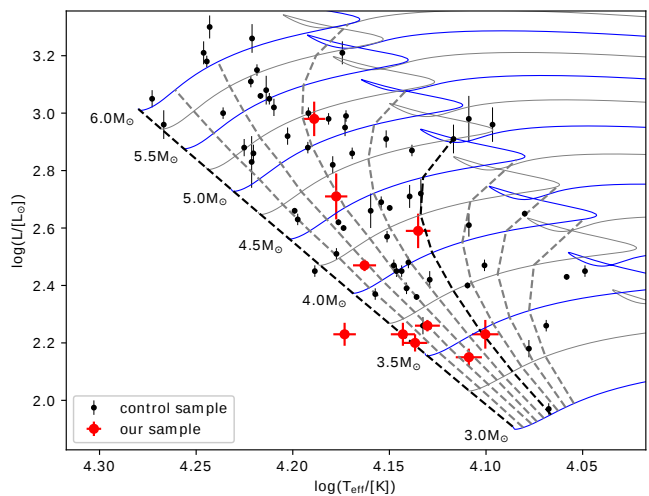}
 \caption{ \LogTeff\ vs. \logl\ diagram for our sample stars (red symbols) and a control sample of well-established SPB stars from the literature (black symbols) with available $uvby\beta$ photometry. Also indicated are isochrones from the ZAMS to 160\,Myr with a spacing of 20\,Myr.}
 \label{fig:hrd}
\end{figure}

\section{\LogTeff versus \logl\ diagram} \label{section_HRD}

To check whether our sample stars fall within the regime of known SPB stars, we collected a control sample of well-established SPB variables (i.e., stars with a detailed asteroseismic analysis available) that also boast $uvby\beta$ photometry \citep{Paunzen15}. We gleaned the $BV$ magnitudes from the compilation by \citet{Kharchenko01}, who compiled an all-sky catalog of more than 2.5 million stars and transformed the available Hipparcos/Tycho $BV$ magnitudes to Johnson $BV$ using a homogenized transformation law. Corresponding $I$ magnitudes were taken from the Hipparcos catalog; $JHK_{\mathrm s}$ magnitudes are from the 2MASS 6X Point Source Working Database \citep{Skrutskie06}.

The final \Teff\ values were derived in two different steps. First, we relied on the calibration by \citet{Napiwotzki93} on the basis of $uvby\beta$ photometry. For the stars without $uvby\beta$ photometry, \Teff\ was derived from the updated relations for MS stars published by \citet{Pecaut13}. A mean value of the calibrated values from $(B-V)_{\mathrm 0}$, $(V-I)_{\mathrm 0}$, and $(V-K_{\mathrm s})_{\mathrm 0}$ was calculated. To test all calibrated values, the VOSA (VO Sed Analyzer) tool v6.0 \citep{Bayo08} was applied to fit the spectral energy distribution (SED) to the available photometry. No outliers were detected, which provides confidence in our results. In addition, a propagation of uncertainties was performed for the photometric observations, from which we deduce a precision of approximately 200\,K for the derived \Teff values.

We calculated the luminosity based on the parallax, apparent magnitude, reddening, and bolometric correction (BC). Parallaxes were taken from Gaia DR2 \citep{Lindegren18}. We interpolated reddening values for stars without any $uvby\beta$ photometry within the corresponding maps published by \citet{Green18} using the distances from \citet{Bailer18}. Employing the derived \Teff\ values, the BC was calculated from the relations listed by \citet{Flower96}. Finally, we used the bolometric magnitude of the Sun (+4.75\,mag) to derive luminosities and errors. For the latter, a complete error propagation was applied. The derived astrophysical parameters of our sample stars are listed in Table \ref{tab:astro}.

Spectral types were taken from the catalog of \citet{Skiff14}, who made a careful assessment of the literature. Almost all classifications originate from the Michigan catalog of two-dimensional spectral types for the HD Stars \citep{Houk99} and are thus based on photographic plates. This likely explains why two stars (HD\,61076 and HD\,168121) are listed with giant luminosities (luminosity class III) but are not located in this region of the HRD. Their spectral types, however, are consistent with the derived \Teff\ values.

In Figure \ref{fig:hrd}, the \logTeff\ versus \logl\ diagram for our sample stars and the control sample of well-established SPB variables is shown. Also indicated are isochrones from the ZAMS to 160\,Myr with a spacing of 20\,Myr, which were calculated with the Modules for Experiments in Stellar Astrophysics (MESA; \citealt{Paxton11, Paxton13, Paxton15, Paxton18}) code. No evolved SPB variables are present in both samples, and there is a noticeable lack of SPB pulsators at or very close to the ZAMS; this fact was already noticed before \cite{Zwintz17} but which is not yet understood. With our work, we add at least two very young SPB stars (HD\,36999 and HD\,61712) to the overall sample. The star HD\,61076 deserves special mention, as it is located way below the ZAMS.

\section{Grid of stellar models} \label{models}
\label{sec:grid_of_models}

We used the MESA and GYRE nonadiabatic oscillation codes \citep{Townsend13, Townsend18} to identify modes of oscillation in our sample stars. An extensive grid of MS stellar models was calculated with four free parameters: initial mass, initial metallicity, age, and overshooting. Initial masses of the stellar models were set within the range 3-5.5\,M$_{\odot}$, with a step-size of $\Delta M$\,=\,0.1\,M$_{\odot}$. Since stellar evolution on the MS does not progress evenly, the time-step $\Delta t$ was parameterized by the maximum allowed rate of change in overall relative hydrogen abundance $dX/X$ = 0.001. Initial metallicities of our stellar models range from 0.003 to 0.03, with a step-size of $\Delta Z$ = 0.003 and additional nodes at Z = 0.019 and 0.02. We also took into account the impact of the overshooting parameter $f_{ov}$ and calculated models with overshooting parameters in the range 0 to 0.04, with a step-size of $\Delta f_{ov}$ = 0.002. In this way, over 85\,000 stellar models were computed. The detailed MESA inlist used in our calculations is given in Appendix \ref{app:mesa}.

As the next step, eigenfrequencies of low-degree g-modes were calculated for each stellar model using the GYRE code. As  a consquence of visibility issues with high-angular degree modes \citep{Aerts10} and computational time limitations, eigenfrequencies were calculated only for low-degree g-modes with $\ell$\,= 1, 2. Furthermore, we decided to restrict our calculations to frequencies above 0.3 c/d for $\ell$\,=\,1 and 0.6 c/d for $\ell$\,=\,2. Below these corresponding frequency thresholds, the frequency spacing of high-order modes is insufficient to prevent misidentification of modes due to uncertainties in the derived stellar parameters. The complete GYRE inlist can also be found in Appendix \ref{app:mesa}. 

Stellar model eigenfrequencies were calculated on nonrotating stellar models, which caused degeneracy in modes with the same angular degree $l$ but different azimuthal order $m$. If necessary, this degeneracy was lifted by introducing rotational splitting for a given mode multiplet, i.e.,
\begin{equation}
    \label{eq:rotational_splitting}
    f_{l, m} = f_{l, 0} + mf_{rot},
\end{equation}{}
where $f_{rot}$ is the rotational frequency of the star.

\section{Mode identification} \label{modes}
Once the grid of stellar models with their g-mode eigenfrequencies was completed, synthetic frequencies of low-degree g-modes calculated with GYRE were compared to the observed frequencies obtained by the frequency analysis described in Section \ref{sub_phot_data_reduction_and_analysis}. In our search algorithm, the accuracy of the fit of a given g-mode combination and the corresponding stellar model was evaluated using the function
\begin{equation}
\label{eq:chi_square}
    \chi^2 = \frac{1}{N}\sum_{i=1}^{N}\frac{(f_{obs, i} - f_{synth, i})^2}{f_{synth, i}},
\end{equation}{}
where $N$ is the total number of detected frequencies viable for cross-match, $f_{obs}$ is the observed frequency, and $f_{synth}$ is its supposed synthetic counterpart.

To reduce the number of stellar models to be checked for each object, the original grid of models was cropped to a subgrid of models with luminosities and effective temperatures within 1$\sigma$ to 3$\sigma$ of our target. To find the best-fit stellar model, we calculated $\chi^2$ for possible combinations of g-mode eigenfrequencies and each stellar model. Finally, the mode combination scoring the highest $\chi^2$ value within a 3\,$\sigma$ error box was selected.

\begin{figure}[t]
 \centering
 \includegraphics[width=0.99\linewidth,keepaspectratio=true]{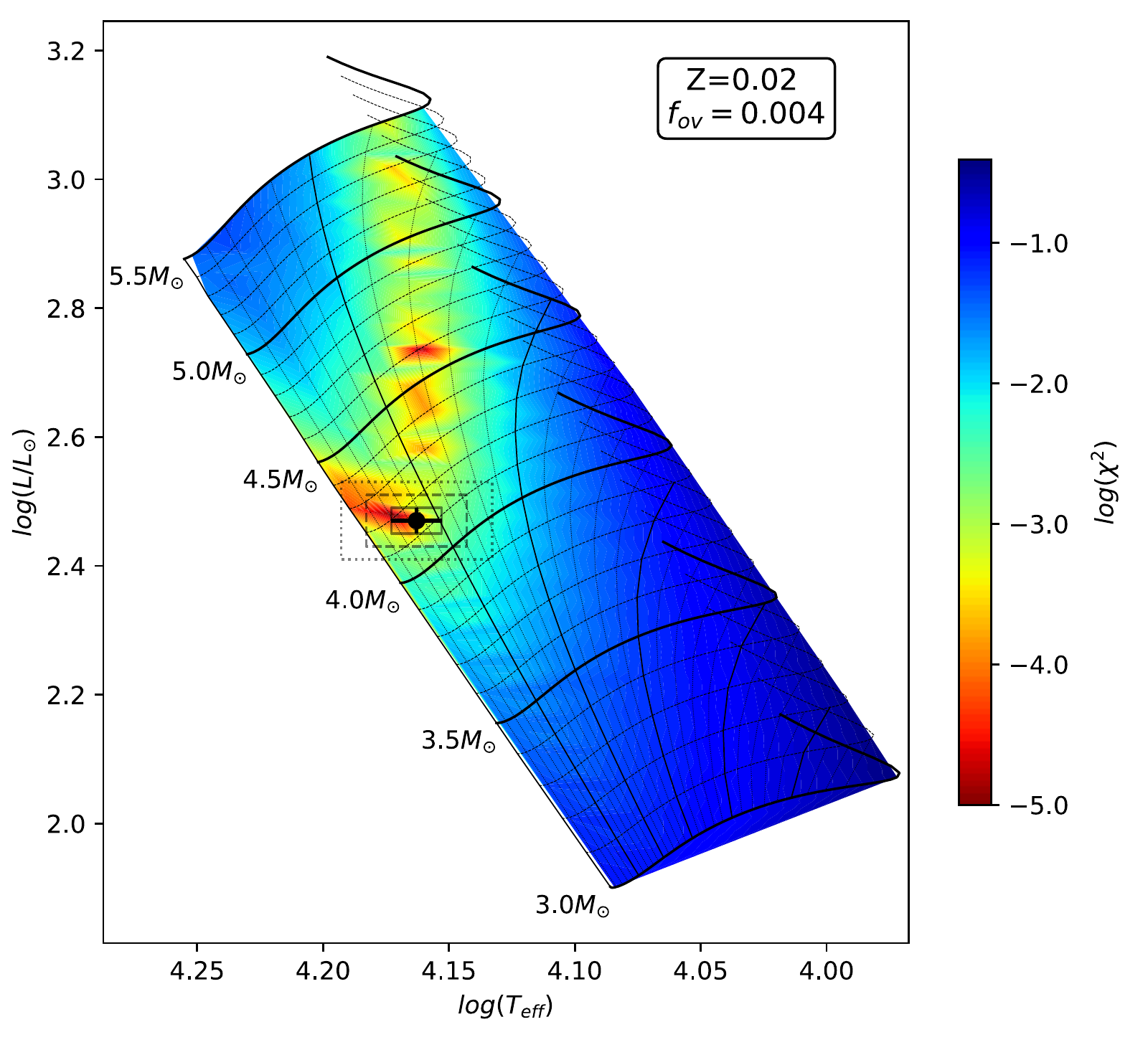}
 \caption{Hertzsprung-Russell diagram (HRD) illustrating a 2D slice of the grid search procedure for the star HD\,48497, using the best-fit mode combination (see Table \ref{tab:modes}). Also shown are evolutionary tracks with initial metallicity and overshooting parameter that provided the best fit of the models to the observed frequencies. The underlying heat map indicates the interpolated $\chi^2$ values on a logarithmic scale. Corresponding isochrones are overlayed in 10\,Myr increments. The location of HD\,48497 is indicated by the black dot, which is surrounded by the corresponding 1$\sigma$, 2$\sigma$, and 3$\sigma$ error boxes.}
 \label{fig:gridsearch}
 
\end{figure}

Initially, eigenfrequencies for non-rotating models were used in the search algorithm. If no satisfactory solution was found, rigid-body rotation was introduced using Eq. \ref{eq:rotational_splitting}. The rotational period was searched for heuristically either by searching for equidistant spacing in the observed frequencies or by assuming that certain couples of frequencies are wings of a dipole triplet where the central peak was not detected. The rotational period was finally derived by a linear fit of Eq. \ref{eq:rotational_splitting} to the detected triplet. A 2D slice of the grid search procedure is shown in Figure \ref{fig:gridsearch}.

Table \ref{tab:modes} lists the detected frequencies and, where applicable, the corresponding modes. In case a frequency multiplet was detected, we also identified the azimuthal order $m$. The rotational periods derived from the frequency spacing of triplet component frequencies are provided in Table \ref{tab:rotation_periods}.

In the case of HD\,168121, the spacing between the different overtones $n$ is relatively small for $\ell$\,=\,1 modes. Furthermore, the three detected frequencies are closely but not symmetrically spaced, which means that they do not belong to an $\ell$\,=\,1 triplet or the triplet is asymmetric because Eq. \ref{eq:rotational_splitting} is no longer valid for fast rotators \citep{Cowling1949ApJ...109..149C}. Another possible explanation is that two frequencies are the product of rotational splitting of the same mode and the third frequency is  a member of a different mode. Because of the lack of any additional frequencies outside the problematic region and our model assuming triplets to be symmetric because of the size of the grid, none of the detected frequencies in this star could be reliably identified.

\begin{table}[t]
\caption{List of all detected frequencies along with the corresponding g-modes characterized by angular degree $\ell$, radial order $n_{\mathrm g}$, and azimuthal order $m$. Modes were extracted from the best-fit mode combination and stellar model using a grid of models described in Section \ref{sec:grid_of_models}. Masses and ages of the stellar models producing the best frequency cross-matches are given in parentheses behind the object identifiers. Age estimation was carried out in cases where models within 0.5 dex of the best-fit model boasted approximately the same age.}
\label{tab:modes}
\footnotesize
\begin{tabular}{llllllll}
\hline
& Frequency & Amplitude & Phase & S/N & $l$ & $n_{\mathrm g}$ & $m$  \\ 
& (c/d) & (mmag) \\
\hline
\multicolumn{5}{c}{HD 36999 (3.6\,M$_\odot$, 25\,Myr)} \\
\hline
$f_{1}$  &     1.50460(2)    &   10.6(9) &      0.41(1)  &  10.6 & 1 & 9  &  +1 \\
$f_{2}$  &     1.43227(2)    &   12.6(8) &      0.66(1)  &  12.6 & 1 & 9  & $-$1 \\
$f_{3}$  &     1.82621(3)    &    5.9(9) &      0.84(3)  &   5.9 & 1 & 7  & 0  \\
$f_{4}$  &     7.44114(3)    &    4.6(9) &      0.98(3)  &   4.3 & 1 & 1  & 0  \\
$f_{5}$  &     2.61151(3)    &    4.9(9) &      0.17(3)  &   4.8 & 1 & 5  & 0  \\
\hline 
\multicolumn{5}{c}{HD 48497 (4.2\,M$_\odot$, 25\,Myr)} \\
\hline
$f_{1}$  &     0.938(2)      &    12(4)  &      0.8(1)   &   7.0 &   &    &   \\
$f_{2}$  &     1.043324(8)   &    12(2)  &      0.68(2)  &   6.6 & 1 & 11 &   \\
$f_{3}$  &     0.89578(6)    &     9(2)  &      0.01(6)  &   5.0 & 1 & 13 &   \\
$f_{4}$  &     1.041(1)      &     8(3)  &      0.2(1)   &   4.5 & 2 & 20 &   \\
$f_{5}$  &     1.265(1)      &     8(2)  &      0.87(5)  &   4.4 & 1 & 9 &   \\
\hline
\multicolumn{5}{c}{HD 61076 (3.2\,M$_\odot$, 2\,Myr)} \\
\hline
$f_{1}$  & 1.0919(3)         &   13(3)   &      0.4(2)   &   4.2 & 1 & 14 &   \\
$f_{2}$  & 2.73251(1)        &   10(1)   &      0.35(2)  &   4.0 & 1 &  5 &   \\
\hline
\multicolumn{5}{c}{HD 61712 (3.6\,M$_\odot$)}\\
\hline
$f_{1}$  & 1.281592(3)       &   18.7(7) &      0.422(7) &  25.4 & 1 & 10 &   \\
$f_{2}$  & 1.513(2)          &    4(1)   &      0.5(2)   &   4.7 & 2 & 15 &   \\
\hline
\multicolumn{5}{c}{HD 66181 (75\,Myr)} \\
\hline
$f_{1}$  & 0.88522(2)        &   14(1)   &      0.54(1)  &  11.5 & 1 & 12 & +1  \\
$f_{2}$  & 0.794(8)          &    9(4)   &      0.4(2)   &   7.7 & 1 & 12 & $-$1 \\
$f_{3}$  & 0.89936(3)        &    8(1)   &      0.91(3)  &   6.6 & 1 & 11 & 0  \\
$f_{4}$  & 0.70470(4)        &    7(1)   &      0.52(3)  &   5.7 & 2 & 26 & 0  \\
$f_{5}$  & 0.84193(3)        &    6(1)   &      0.54(3)  &   4.5 & 1 & 12 & 0  \\
\hline
\multicolumn{5}{c}{HD 86424 (3.5\,M$_\odot$, 50\,Myr)} \\
\hline
$f_{1}$  & 1.084518(3)       &   12.7(3) &      0.544(3) &  13.2 & 1 & 12 &    \\
$f_{2}$  & 0.10747(1)$^{a}$  &    4.4(2) &      0.873(9) &   4.5 &   &    &    \\
$f_{3}$  & 0.95053(4)        &    4.9(3) &      0.50(1)  &   5.1 & 1 & 14 &    \\
\hline
\multicolumn{5}{c}{HD 97895 (4.8\,M$_\odot$, 50\,Myr)} \\
\hline
$f_{1}$  & 0.922626(7)       &   16.1(3) &      0.667(3) &   7.4 & 1 & 12 & 0  \\
$f_{2}$  & 0.861588(8)       &   15.5(3) &      0.471(3) &   7.6 & 1 & 12 & $-$1 \\
$f_{3}$  & 0.98459(1)        &   10.1(3) &      0.275(5) &   4.2 & 1 & 12 & +1  \\
$f_{4}$  & 1.82168(1)        &    9.3(3) &      0.611(5) &   4.3 & 1 &  6 & 0  \\
\hline
\multicolumn{5}{c}{HD 115067} \\
\hline
$f_{1}$  & 0.62165(2)        &  10(1)    &      0.15(2)  &   7.4 & 1 & >18 &    \\
$f_{2}$  & 0.18665(2)$^{a}$  &   7(1)    &      0.27(2)  &   4.9 &   &    &    \\
\hline
\multicolumn{5}{c}{HD 163285 (3.8\,M$_\odot$, 40\,Myr)} \\
\hline
$f_{1}$  & 0.82863(1)        & 17.0(8)   &      0.319(8) &   7.1 & 1 & 15 &    \\
$f_{2}$  & 1.07449(1)        & 13.3(8)   &      0.22(1)  &   5.6 & 1 & 11 &    \\
$f_{3}$  & 1.20341(1)        & 11.6(8)   &      0.94(1)  &   4.9 & 2 & 18 &    \\
\hline
\multicolumn{5}{c}{HD 168121} \\
\hline
$f_{1}$  & 0.75915(2)        & 11(1)     &      0.30(2)  &   4.9 &   &    &    \\
$f_{2}$  & 0.79316(2)        & 10(1)     &      0.99(2)  &   4.4 &   &    &    \\
$f_{3}$  & 0.77929(2)        &  9(1)     &      0.06(2)  &   4.1 &   &    &    \\
\hline
\end{tabular}
\newline
$^{a}$ Frequency overlap of synthetic modes too severe to prevent misidentification
\end{table}

\begin{table}[t]
\caption{List of rotation periods for objects with (partially) detected $\ell$\,=\,1 triplets, based on Eq. \ref{eq:rotational_splitting}.}
\label{tab:rotation_periods}
\centering
\begin{tabular}{l c }
\hline
Object & Period(d) \\
\hline
HD 36999  & 27.7  \\
HD 66181  & 21.9(6)  \\
HD 97895  & 16.26(7)  \\
\hline
\end{tabular}
\end{table}

\section{Results} \label{results}

In the following, the results on the individual stars are discussed. Figures \ref{fig:best_fit_mode_combination} and \ref{fig:best_fit_mode_combination2} provide a discussion of the properties of the best-fit stellar models.

{\it HD\,36999:} This star, which is listed as a young stellar object (a star in its earliest stages of development, i.e., either a protostar or a PMS) in SIMBAD \citep{SIMBAD}, is a member of the Orion OB association \citep{Tian96} with an estimated age of 2.2\,Myr \citep{Wolff04}, which is well in line with our results. This makes HD\,36999 a particular interesting object because it is located on, or slightly below, the ZAMS. As has been pointed out (cf. Section \ref{introduction}), only very few SPB stars of comparable age are known so far \citep{Gruber12,Zwintz17}. 

To investigate the evolutionary status of this object, we checked the available LAMOST DR5 spectrum \citep{lamost1,lamost2} for the presence of emission lines or other telltale signs of PMS stars. Interestingly, the spectrum clearly shows numerous strong emission lines, for instance, \ion{O}{II}\,3728\,\AA, H$\beta$\,4863\,\AA, \ion{O}{III}\,4960\,\AA, \ion{O}{III}\,5007\,\AA, H$\alpha$\,6563\,\AA,\ and \ion{S}{II}\,6716,6731\,\AA. This is in agreement with the emission profile of the nebula that pervades the corresponding sky region and apparently contaminates the LAMOST spectrum of HD\,36999 and the spectra of numerous other stars in the vicinity. Contamination of LAMOST spectra by \ion{H}{II} regions is a known issue \citep{Hou16}. We assume that this may be caused by fiber drift, issues with the pipeline software responsible for the background substraction, or by sampling the background during the integration while the seeing is excellent.

Because of the higher luminosity derived from Gaia DR2 data, we determined that the best-fit model boasts a mass of 3.6\,$M_{\odot}$ and an age of 25\,Myr; these values are somewhat higher than those given by \citet{Wolff04} of 3.26\,$M_{\odot}$ and 2.2\,Myr. Our analysis suggests a rather low value of $f_{ov}$ < 0.01. We achieved best results by regarding frequencies $f_1$ and $f_2$ as side lobes of a rotationally split triplet. Owing to the non-detection of the central peak of this supposed triplet and underestimated frequency uncertainties provided by PERIOD04, we were unable to properly estimate the uncertainty of the rotation period.

{\it HD\,48497:} \citet{Niemczura09} presented a detailed abundance analysis of this star, which has a very low \vsini\ value of 13\,\kms\ and underabundances of Cr, Sr, and Ni of about 0.5\,dex as compared to the solar values. Therefore, \citet{Yushchenko15} include HD\,48497 in their list of possible $\lambda$ Bootis star candidates. These stars constitute a small group of A- to F-type stars characterized by the depletion of Fe-peak elements of up to 2\,dex \citep{Murphy17}. However, HD\,48497 is too hot and does not show the typical elemental abundance pattern of $\lambda$ Bootis stars. It therefore cannot be considered a member of this group. From our analysis, we find that the models with a mass of 4.2\,$M_{\odot}$ and an age of 25\,Myr provide the best fit to the observed frequencies. It is noteworthy that during the analysis, we discovered that frequency $f_1$ is  a side lobe of the 1\,d alias frequency that is ubiquitous in ground-based observations. Consequently, this frequency was disregarded during the grid search procedure.

{\it HD\,61076:} The available Hipparcos parallax of 0.96(78)\,mas led to the conclusion that this star is a giant with a luminosity of 910\,L$_\odot$ \citep{Hohle10}. The Gaia DR2 parallax, on the other hand, places this star well below the ZAMS. We find that, for the given luminosity, a \Teff\ difference of 1\,000\,K is needed to shift HD\,61076 to the ZAMS for models with a metallicity of Z = 0.02. The published spectral type of B5/7 might indicate some peculiarities that render the temperature estimation uncertain. We strongly suspect that the \Teff\ determination based on the Gaia DR2 parallax is erroneous; however, only further photometric or spectroscopic data can shed more light on this issue and the nature of this star. 

Since our grid consists of stellar models with a wide range of metallicity and overshooting parameter values, we performed a grid search for the best-fit g-mode combination despite the apparently peculiar nature of this star. As expected, low-metallicity models with Z < 0.01 were able to shift the ZAMS sufficiently toward the position of our object in the HRD. The best-fit model indicates a 3.2\,$M_{\odot}$ object near the ZAMS (2\,Myr) with an initially low metallicity of Z = 0.009 and the rather high value of $f_{ov} > 0.03$. However, the high value of $f_{ov}$, the age of the object, and its peculiar location in the HRD cast doubt on the reliability of $f_{ov}$. Lack of time on the ZAMS should prevent $f_{ov}$ from having any significant impact on the internal structure of the star. In addition, if we exclude models outside the 1$\sigma$ confidence interval, the previously stated constraint on $f_{ov}$ vanishes (see Fig. \ref{fig:best_fit_mode_combination}). Despite these uncertainties, the scarcity of very young SPB stars and the apparently peculiar nature of this object warrant a closer look at HD\,61076 in future studies.

{\it HD\,61712:} \citet{Mannings98} presented evidence that this star might be a PMS star with significant IR excess. An inspection of the SED of the object with the VOSA tool clearly confirms the presence of IR excess. However, a PMS nature cannot be unambiguously established by this criterion alone because similar excesses have been observed in stars that have already reached the MS \citep{Montesinos09}. One definite diagnostic criterion would be the presence of emission lines; however, to the best of our knowledge, no spectrum of this object is available. Our analysis failed to provide an unambiguous age estimation because multiple stellar models with similar values of $\chi^2$ were obtained for the best-fit g-mode combination. However, these models uniformly point to a mass of 3.6\,$M_{\odot}$. In summary, HD\,61712 is a candidate young SPB star worthy of further attention.

{\it HD\,66181:} No detailed investigations of this object are available in the literature. Unfortunately, although the best-fit g-mode combination was found, we were unable to put constraints on the mass of the object. However, from the detection of a dipole triplet, the rotational period could be calculated (see Table \ref{tab:rotation_periods}).

{\it HD\,86424:} This star is a visual binary with a magnitude difference of the components (HD 86424 and CD$-$41\,5479B) of 2.3\,mag and a separation of 9$\farcs$7 \citep{Sinachopoulos88}. Using the available Gaia DR2 parallaxes, \citet{Bailer18} estimate distances between 547 and 573\,pc for HD 86424 and 585 and 607\,pc for CD$-$41\,5479B. The two stars, therefore, do not form a physical system. Assuming an identical reddening value and the given distance, we derive an absolute magnitude of about +2.5\,mag for the fainter component. This is typical for a F0\,V star, which is also compatible with the available optical and near-infrared  colors. In this spectral region is situated the blue border of the $\gamma$ Doradus instability strip \citep{Bradley15}, which is populated by g-mode pulsators with periods between 0.5 to 5\,d. We are not able to rule out that CD$-$41\,5479B is  the source of the detected variability because both stars are covered by the apertures of the employed data sources. The parameters of the two best-fit models point to a mass of 2.5\,$M_{\odot}$ and an approximate age of 50\,Myr. However, it has to be pointed out that the best-fit stellar models within the 1$\sigma$ confidence interval boast a rather high metallicity of Z = 0.027. As we do not have any additional information about the metallicity of HD 86424, we are unable to confirm the validity of our models.

{\it HD\,97895:} With a Galactic latitude of $+$29$\degr$ and low radial velocity \citep{Kordopatis13}, this star is an unusual hot B-type star because it is situated most certainly in the thick disk and is characterized by a metallicity significantly different from the solar value \citep{Miranda16}. The SPB stars of such metallicities are valuable testbeds for the calibration of pulsational models \citep{Miglio07}. The observed frequencies can be closely matched by dipole modes on 50\,Myr old 4.8\,$M_{\odot}$ stellar models. From the rotational splitting of one of the observed frequencies, we were able to determine the rotational period. Despite our expectations, our best-fit models do not indicate metallicities significantly different from the solar value. Models with a low value of $f_{ov}$ produce a more accurate fit to the observed frequencies.

{\it HD\,115067:} With an age of about 90\,Myr, this is the most evolved star in our sample. Otherwise, no detailed investigations are available in the literature. Unfortunately, in the case of this object, only one frequency suitable for cross-identification was available, which is insufficient to perform a grid search. Table \ref{tab:modes} contains a lower estimate on the radial order of the observed frequency $f_1$ under the assumption that it corresponds to a dipole mode.

{\it HD\,163285:} No detailed investigations of this object are available in the literature. High-metallicity stellar models with a mass of 3.8\,$M_{\odot}$ and an age of 40\, Myr provide the best fit to the observed frequencies. The value $f_{ov}$ can be constrained within the range from 0.024 to 0.030.
{\it HD\,168121:} No detailed investigations of this object are available in the literature.

\section{Conclusions} \label{conclusion}

We collected and analyzed extensive sets of photometric time series data of ten hitherto unidentified SPB stars with the aim of describing their pulsational properties and identifying pulsational modes. Astrophysical parameters were calculated and the location of our sample stars in the \logTeff\ versus \logl\ diagram was investigated. We calculated current pulsational models to identify pulsational modes in our sample stars. An extensive grid of stellar models and its corresponding eigenfrequencies were calculated.

For eight objects, the observed frequencies were successfully cross-matched with a best-fit g-mode combination using our grid search algorithm. From the best-fit stellar model, we were able to constrain the astrophysical properties of our sample stars. In the case of five objects (HD~48497, HD~61076, HD~61712, HD~86424, and HD~163285), we were able to constrain masses down to the resolution limit of the grid, i.e., 0.1\,$M_{\odot}$. Employing the ages of the corresponding best-fit models (see Table \ref{tab:modes}), we were also able to derive information on the age of seven stars. HD\,36999 is a particular interesting object because it is located on, or slightly below, the ZAMS. HD\,61712 is another candidate young SPB star. A special case is HD~61076, which may be a very young and low-metallicity object, according
to the best-fit models. On the opposite side, the best-fit models for HD 86424 and HD 163285 indicate a higher than average metallicity of 0.027. In accordance with our expectations, no evolved SPB stars are present in our sample.

As far as overshooting parameter $f_{ov}$ is concerned, our results can be divided into three groups. The first group consists of HD~48497, HD 61712, HD~66181, and HD~86424, and boasts a wide range of best-fit values of $f_{ov}$, which demonstrates the need to better constrain this parameter using g-mode pulsations in stars with convective cores. In the second group of objects (HD~36999 and HD~97895), there is a clear tendency for models with low values of $f_{ov}$ to produce better frequency cross-matches. Finally, in the third group of objects (HD~61076 and HD~163285), the best-fit models confined $f_{ov}$ to intervals above 0.02 (see Figures \ref{fig:best_fit_mode_combination} and \ref{fig:best_fit_mode_combination2}).

With the present study, we significantly enlarge the sample of known SPB stars with reliable pulsational mode identifications. We furthermore demonstrate the feasibility of using ground-based observations to perform basic asteroseismological analyses of SPB stars. While our results do not reach the accuracy of previous studies based exclusively on space photometry (e.g., \citealt{Szewczuk18}), they nevertheless constitute a significant improvement on the constraints provided by the uncertainties in $T_{eff}$ and luminosity derived from photometry and Gaia data. In theory, the presented approach can also be used with space-based observations. However, owing to the much higher number of frequencies detected in these data, the grid search algorithm should be replaced by a more refined and less computationally heavy cross-matching algorithm. Such effort may lead to much tighter constraints on stellar parameters, which will help to shed more light on the internal processes at work in upper MS stars.

\section*{Acknowledgments}
The research of M.F. was supported by the Slovak Research and Development Agency under the contract No. APVV-15-0458 and internal grant No. VVGS-PF-2018-758 of the Faculty of Science, P. J. {\v S}af{\'a}rik University in Ko{\v s}ice. This work presents results from the European Space Agency (ESA) space mission Gaia. Gaia data are being processed by the Gaia Data Processing and Analysis Consortium (DPAC). Funding for the DPAC is provided by national institutions, in particular the institutions participating in the Gaia MultiLateral Agreement (MLA). The Gaia mission website is https://www.cosmos.esa.int/gaia. The Gaia archive website is https://archives.esac.esa.int/gaia. This research has made use of the SIMBAD database and the VizieR catalog access tool, operated at CDS, Strasbourg, France.
\bibliographystyle{aa}
\bibliography{SPBs}

\begin{appendix}
\onecolumn
\section{Amplitude spectra}
\label{app:frequencies}

\begin{figure*}[h!]

 \includegraphics[width=0.48\linewidth]{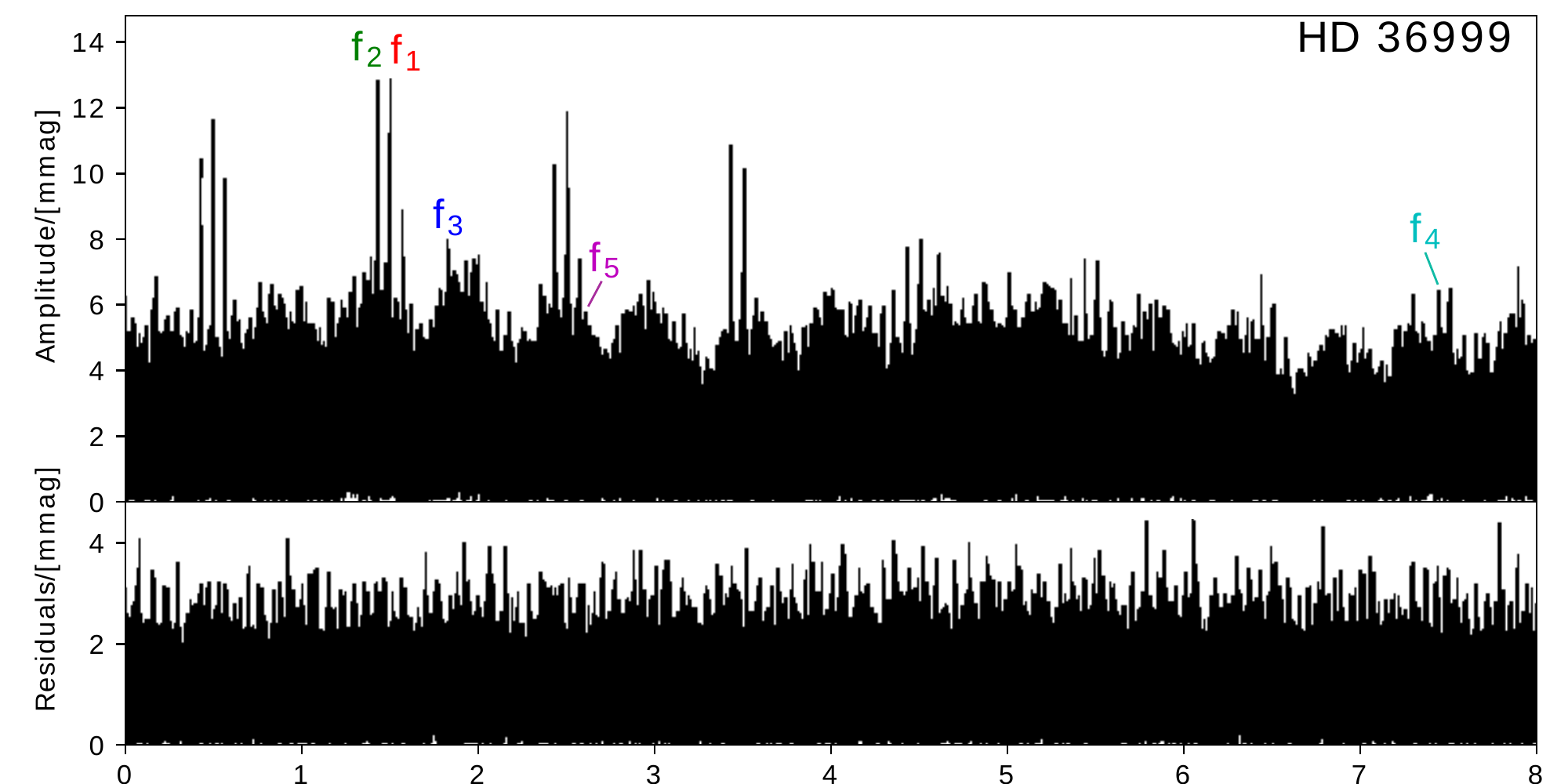}
 \includegraphics[width=0.48\linewidth]{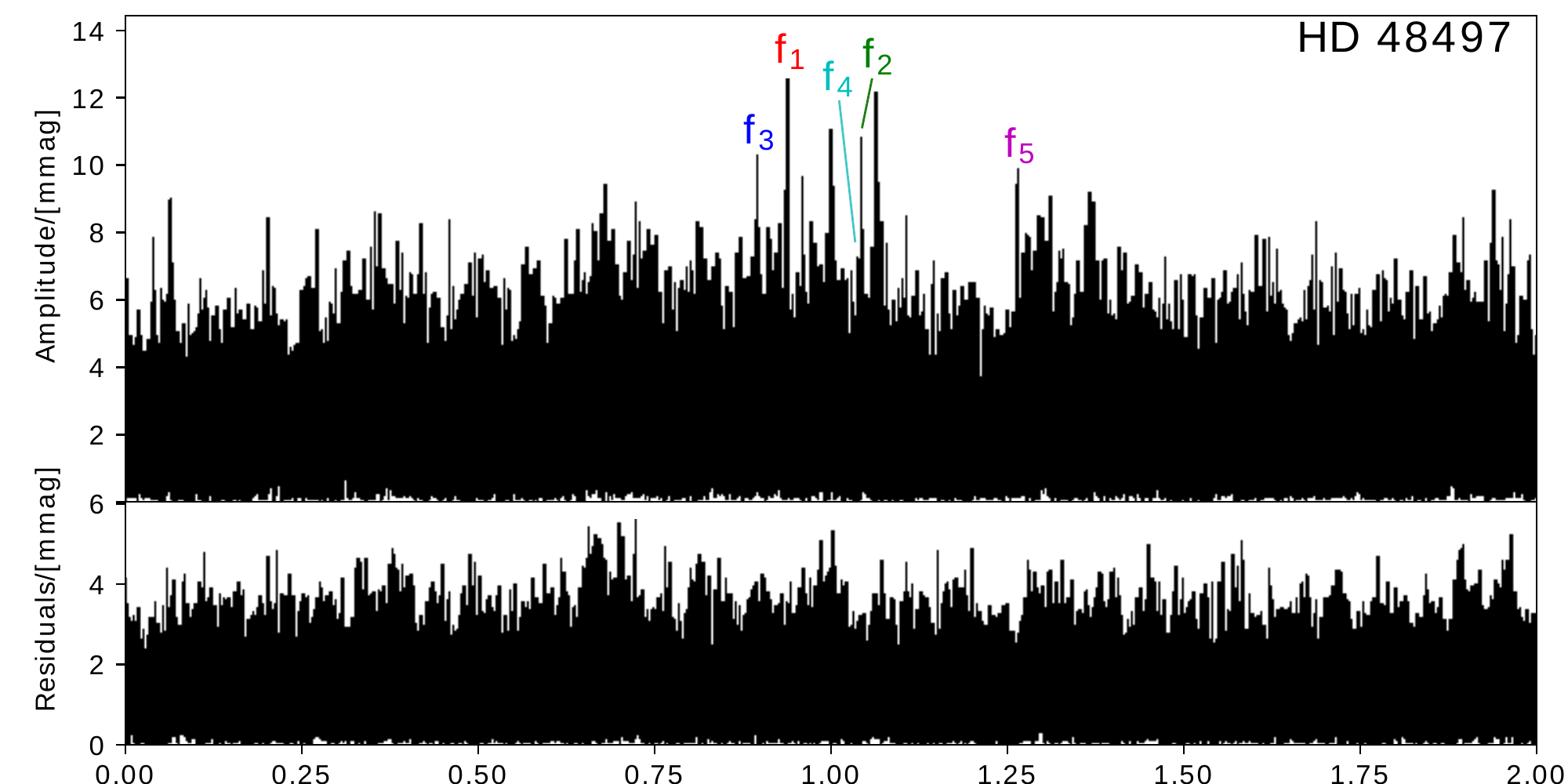}
 
 \includegraphics[width=0.48\linewidth]{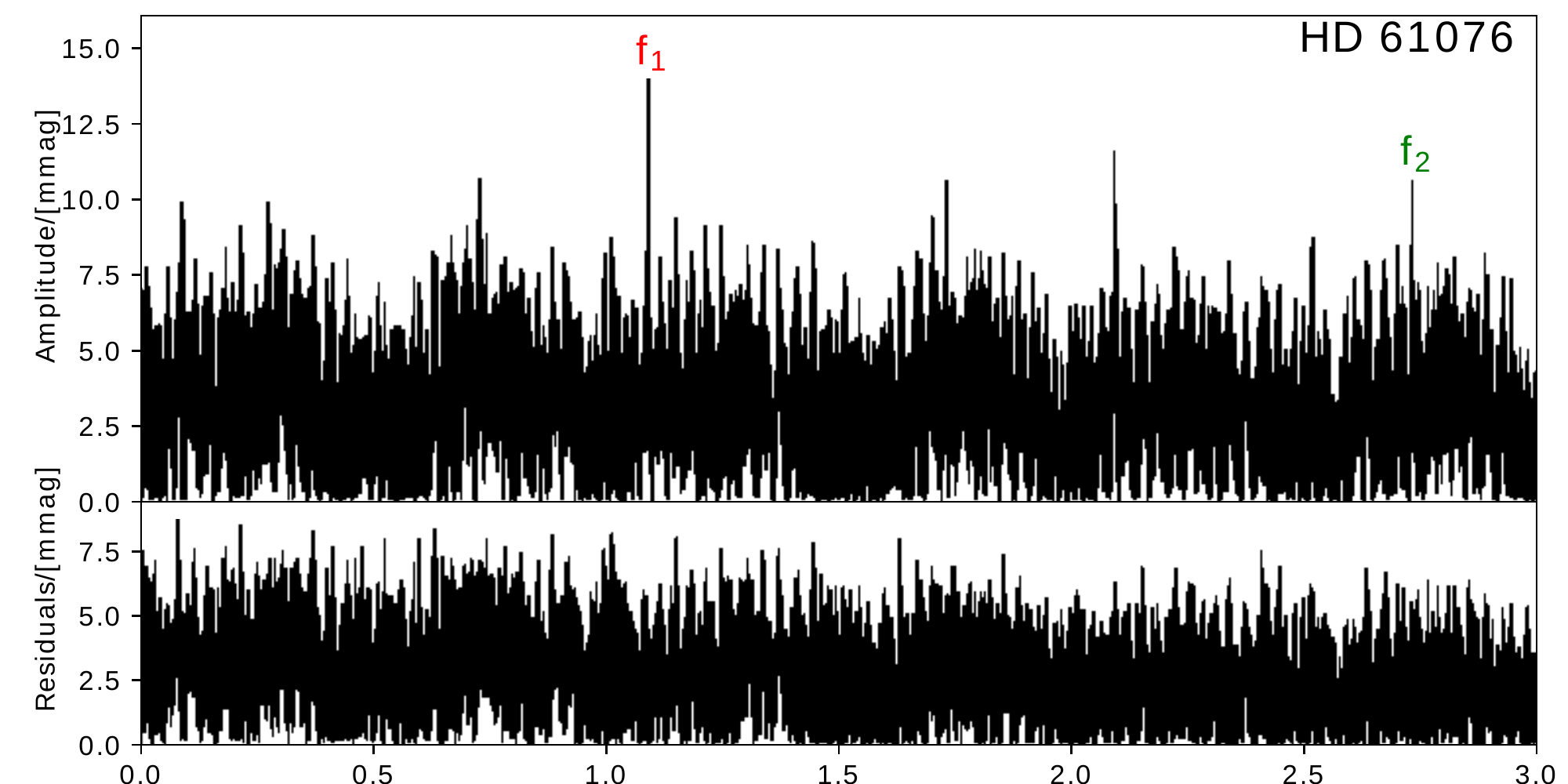}
 \includegraphics[width=0.48\linewidth]{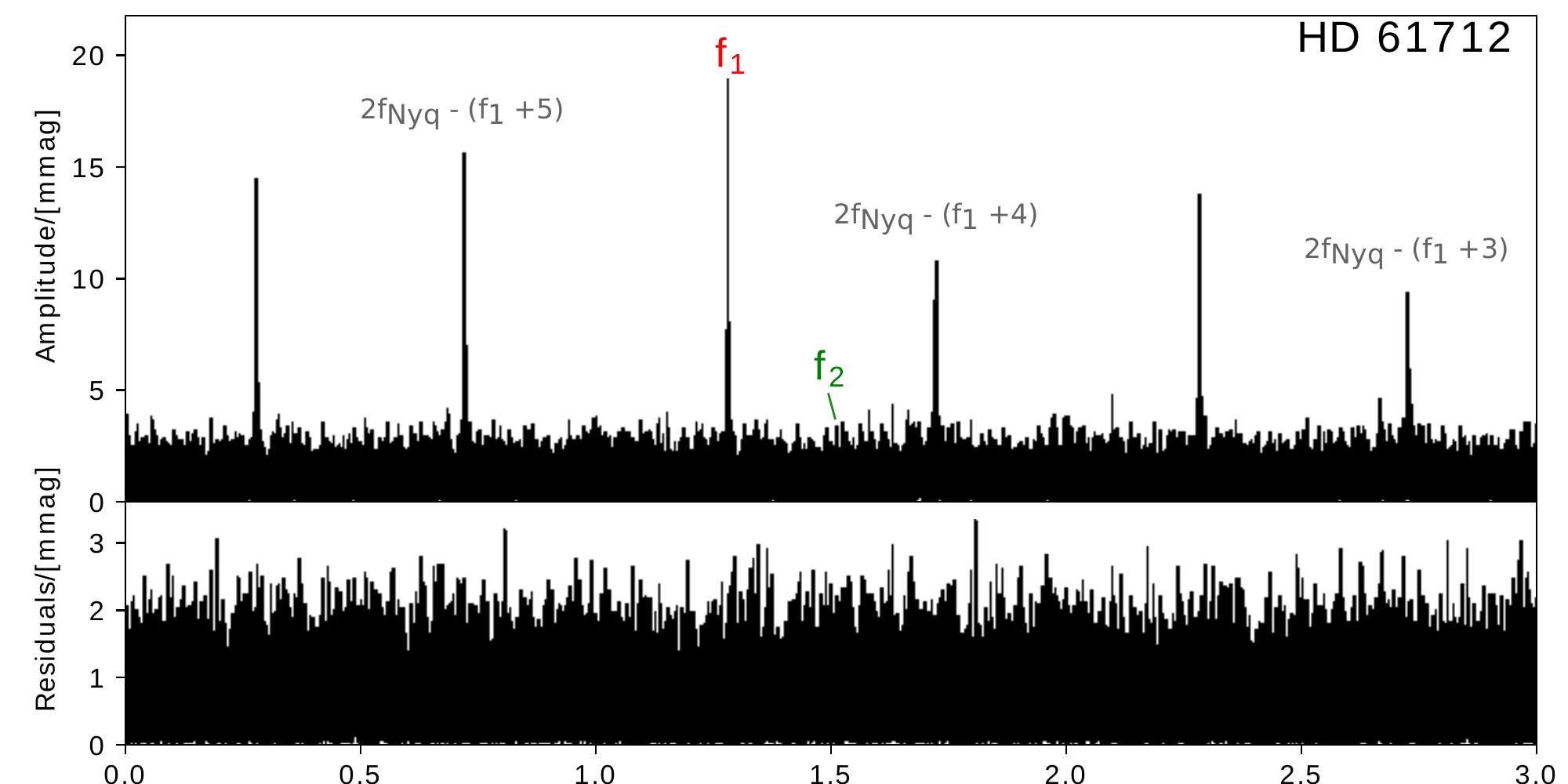}
 
 \includegraphics[width=0.48\linewidth]{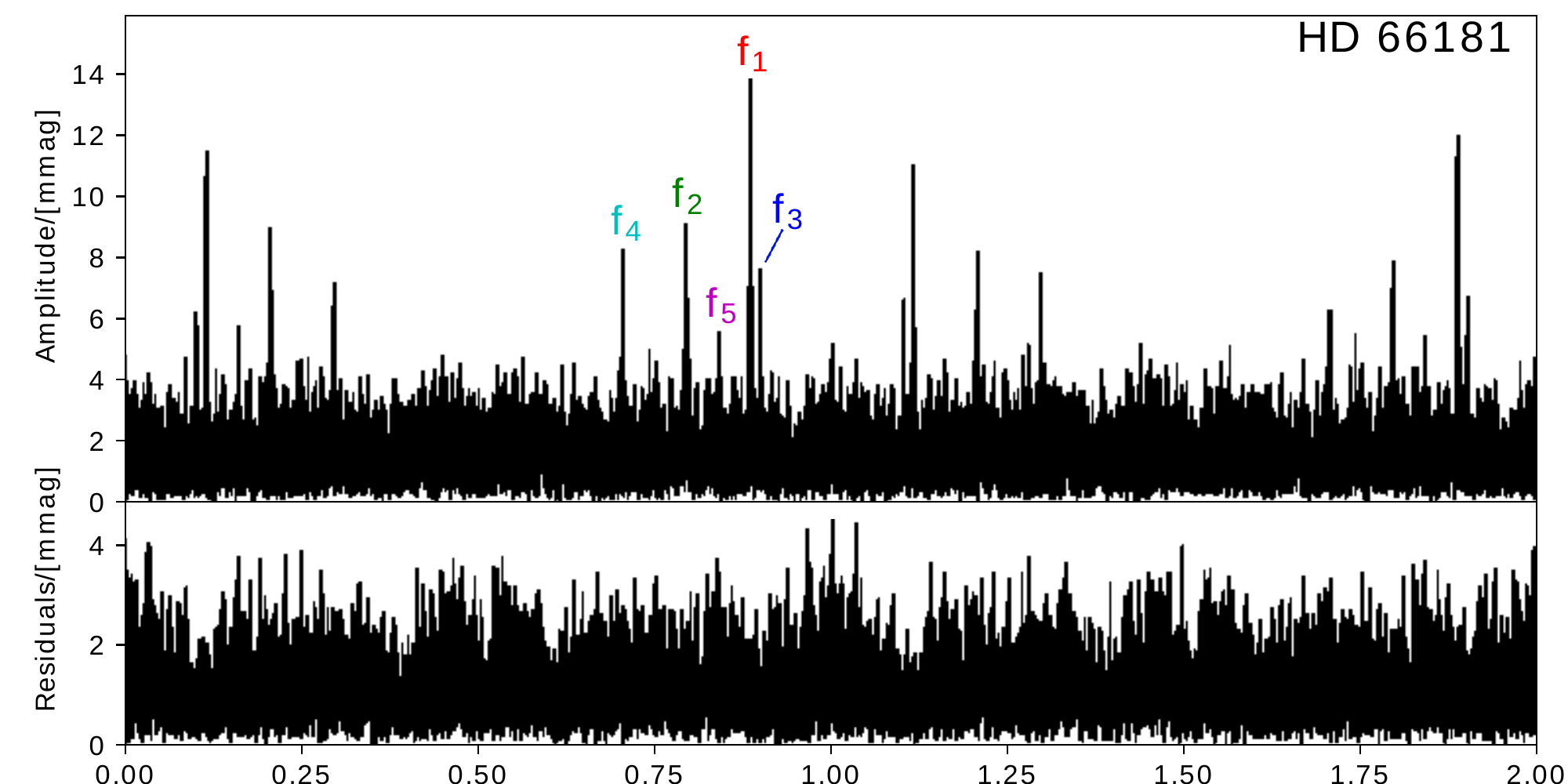}
 \includegraphics[width=0.48\linewidth]{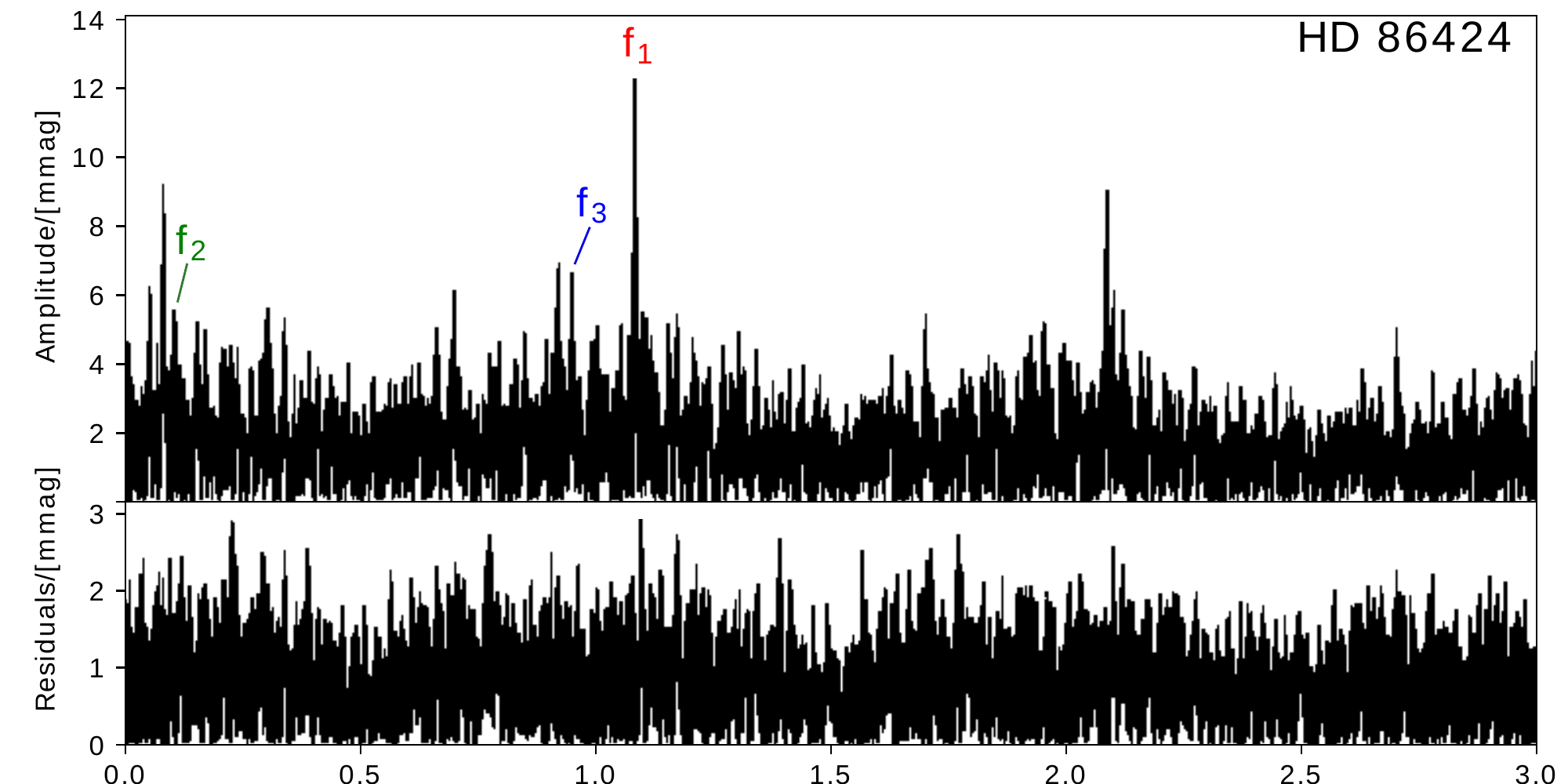}
 
 \includegraphics[width=0.48\linewidth]{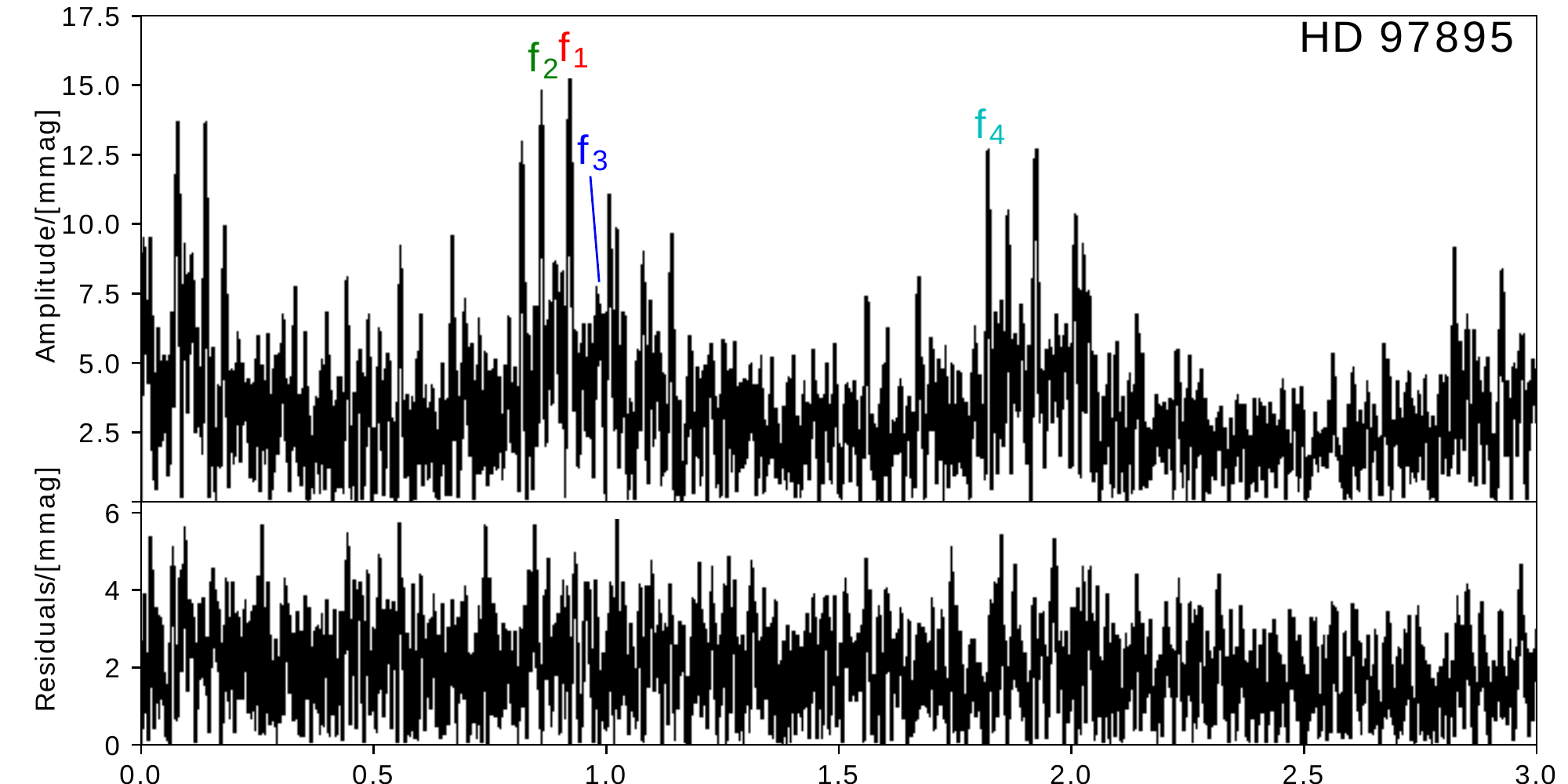}
 \includegraphics[width=0.48\linewidth]{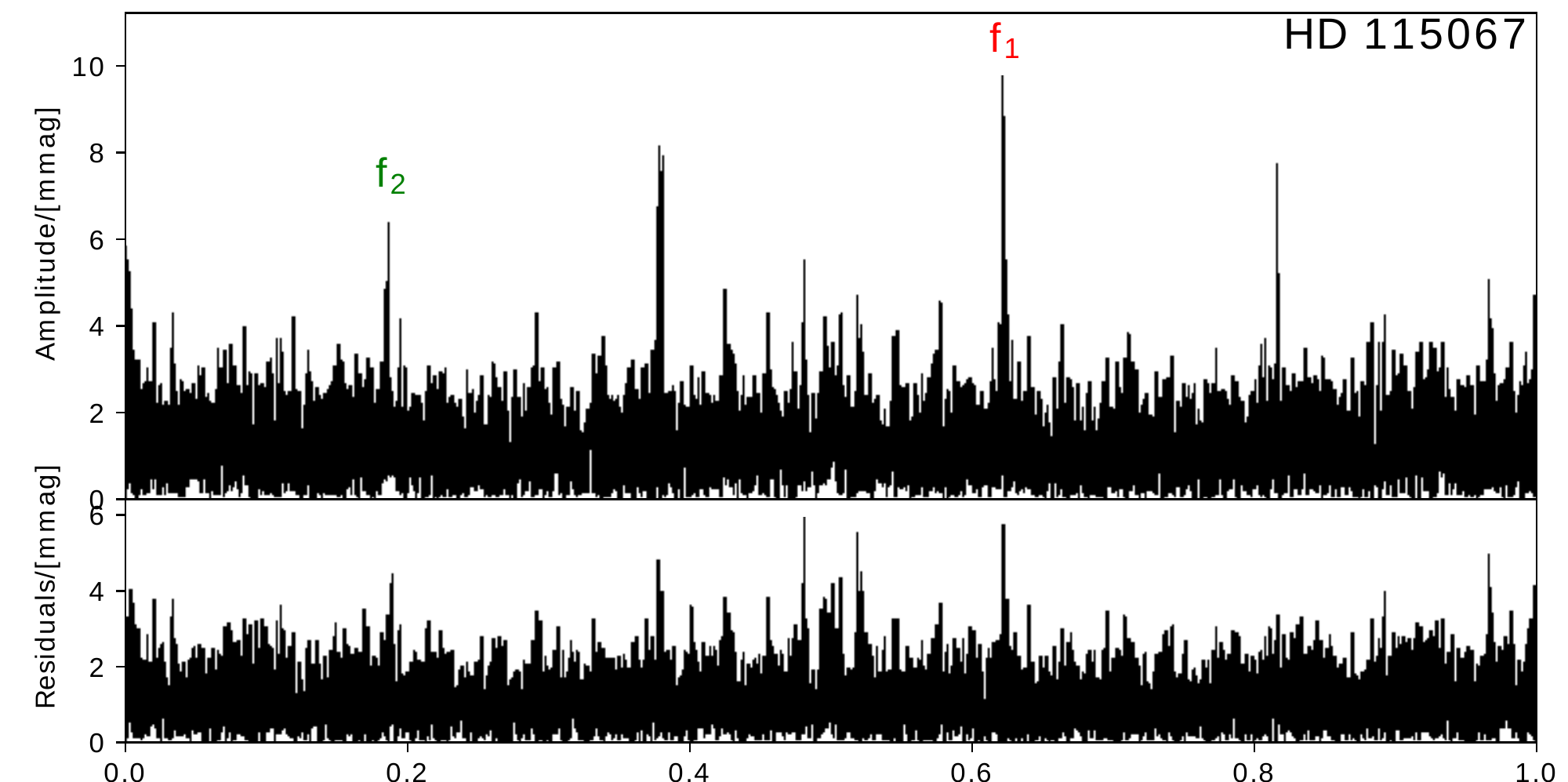}
 
 \includegraphics[width=0.48\linewidth]{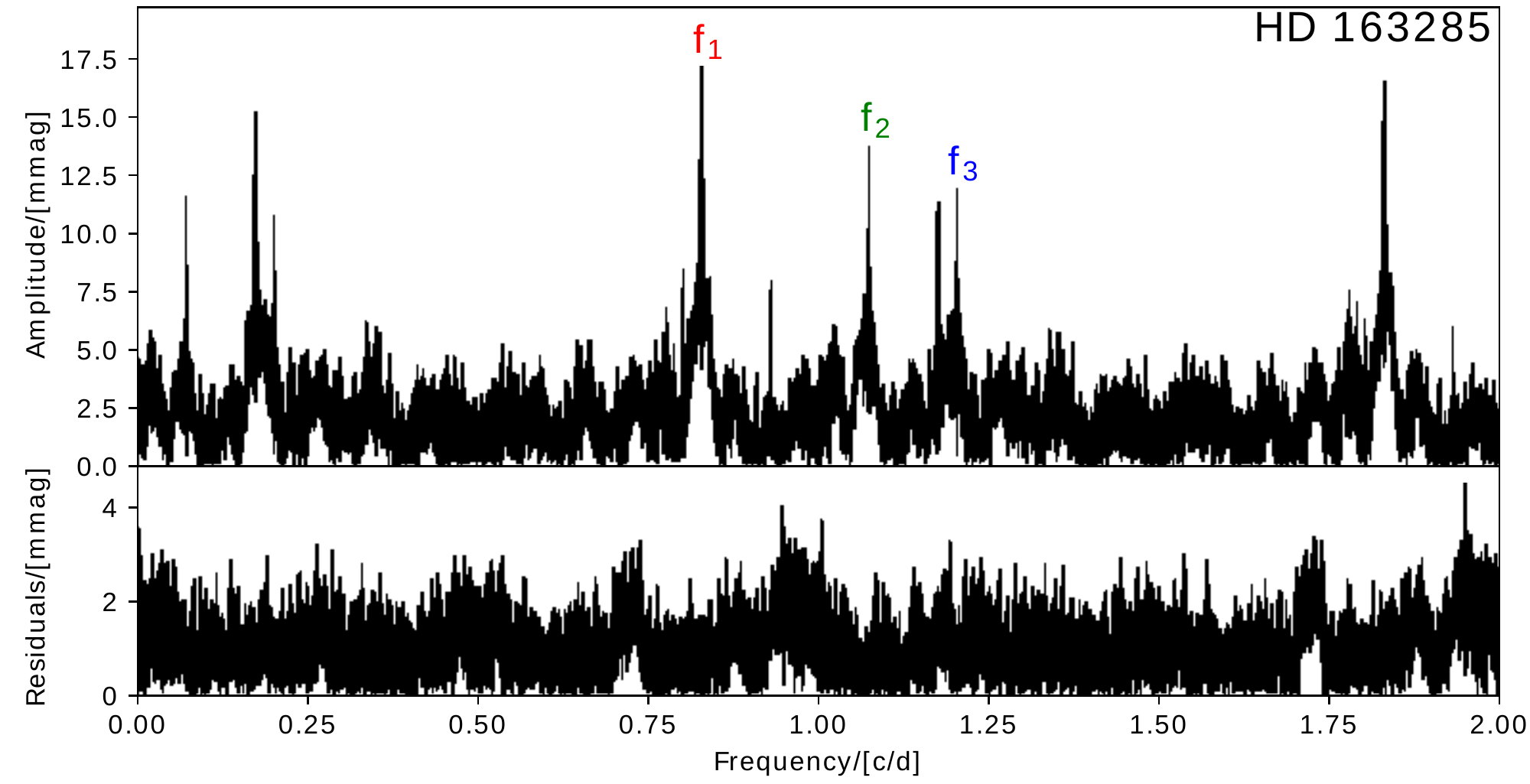}
 \includegraphics[width=0.48\linewidth]{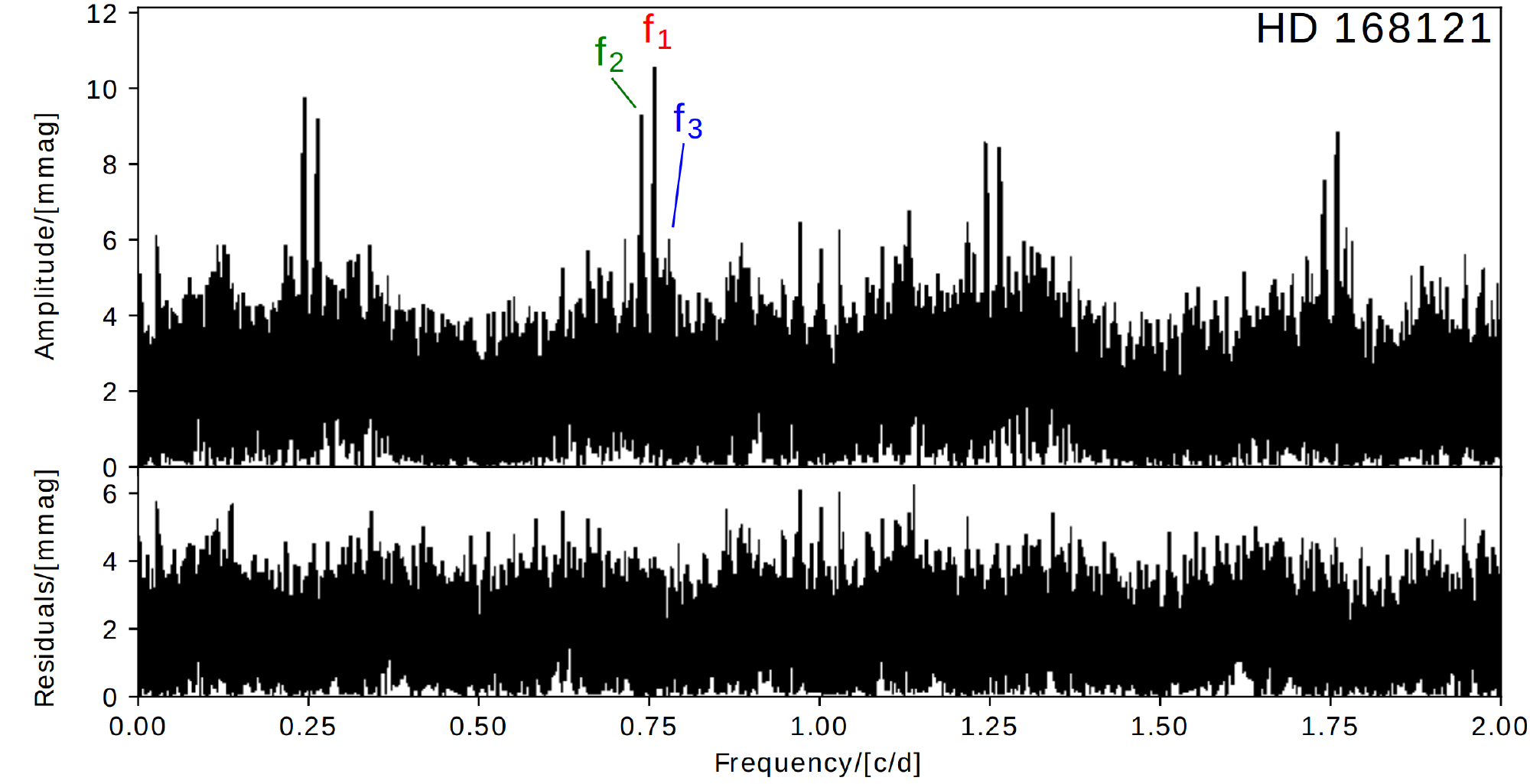}
 
 \caption{Results of the frequency analysis of the light curves of our sample stars. The top panels of each plot illustrate the original amplitude spectra. Significant frequencies, as listed in Table \ref{tab:modes}, are identified. The bottom panels shows the residuals after subtracting the indicated frequencies and corresponding aliases.}
         \label{fig:amplitude_spectra}
\end{figure*}
\twocolumn
\section{MESA and Gyre inlists}
\label{app:mesa}
The grid of stellar models was produced using the following MESA inlist file. Empty parameter values were filled sequentially for each node of the grid.

\begin{verbatim}
&star_job
    create_pre_main_sequence_model = .false.
    write_profile_when_terminate = .false.
    show_log_description_at_start = .false.
    change_lnPgas_flag = .true.
    change_initial_lnPgas_flag = .true.
    new_lnPgas_flag = .true.
    change_Z = .true.
    change_initial_Z = .true.
    new_z =
/ !end of star_job namelist

&controls
    write_profiles_flag = .true.
    profile_interval = 15
    history_interval = 3
    write_header_frequency = 999999
    terminal_interval = 999999
    photo_interval = 5000000
    log_directory =
    
    initial_mass =
    mixing_length_alpha = 1.73
    set_min_D_mix = .true.
    min_D_mix = 0
    
    overshoot_f_above_burn_h_core =
    overshoot_f0_above_burn_h_core = 0.005
    dX_div_X_hard_limit = 1d-3
    
    delta_lg_XH_cntr_max = -1
    delta_lg_XH_cntr_limit = 0.05
    
    alpha_semiconvection = 0.01
    write_pulse_data_with_profile = .true.
    pulse_data_format = "GYRE"
    add_center_point_to_pulse_data = .true.  
    add_double_points_to_pulse_data = .true.
    calculate_Brunt_N2 = .true.
    xa_central_lower_limit_species(1) = "h1"
    xa_central_lower_limit(1) = 1d-3
    
    when_to_stop_rtol = 1d-3
    when_to_stop_atol = 1d-3
    
    mixing_D_limit_for_log = 1d-4
    use_Ledoux_criterion = .true.
    D_mix_ov_limit = 0d0
    which_atm_option = "photosphere_tables"
    cubic_interpolation_in_Z = .true.
    
    ! test with "Gold Standard" tolerances
    threshold_grad_mu_for_double_point = 10.
    newton_iterations_limit = 20
    max_tries = 20
    iter_for_resid_tol2 = 30
    tol_residual_norm1 = 1d-9
    tol_max_residual1 = 1d-7
    tol_correction_norm = 1d-9
    tol_max_correction = 1d-7
/ ! end of controls namelist
\end{verbatim}

After the grid of stellar models was created, each stellar model was then used to calculate the frequencies of dipole and quadrupole g-modes using the following GYRE inlist:

\begin{verbatim}
/ 
&model
model_type = 'EVOL' 
file = 
file_format = 'MESA'
uniform_rot = .TRUE.
Omega_units = 'CYC_PER_DAY'
Omega_rot = 0.0
repair_As = .TRUE.
/
&constants
/
&mode
l = 1
n_pg_min = -25
tag = 'l=1'
/
&mode
l = 2
n_pg_min = -30
tag = 'l=2'
/
&osc
nonadiabatic = .TRUE.
/
&num
diff_scheme = 'COLLOC_GL4'
/
&scan
grid_type = 'INVERSE'
freq_min_units = 'CYC_PER_DAY'
freq_max_units = 'CYC_PER_DAY'
freq_min = 0.321
freq_max = 10.0
n_freq = 100
tag_list = 'l=1'
/
&scan
grid_type = 'INVERSE'
freq_min_units = 'CYC_PER_DAY'
freq_max_units = 'CYC_PER_DAY'
freq_min = 0.6
freq_max = 10.0
n_freq = 100
tag_list = 'l=2'
/
&grid
alpha_osc = 10
alpha_exp = 2
n_inner = 5
/
&nad_output
summary_file = 
freq_units = 'CYC_PER_DAY'
summary_file_format = 'TXT'
summary_item_list = 'M_star,R_star,l,m,n_pg,
n_p,n_g,freq,freq_units,E,E_norm,eta'
/   
\end{verbatim}

\onecolumn
\section{$\chi^2$ distributions for the best-fit g-mode combinations}
\label{app:mode_identification}

\begin{figure*}[h!]

 \includegraphics[width=0.48\linewidth]{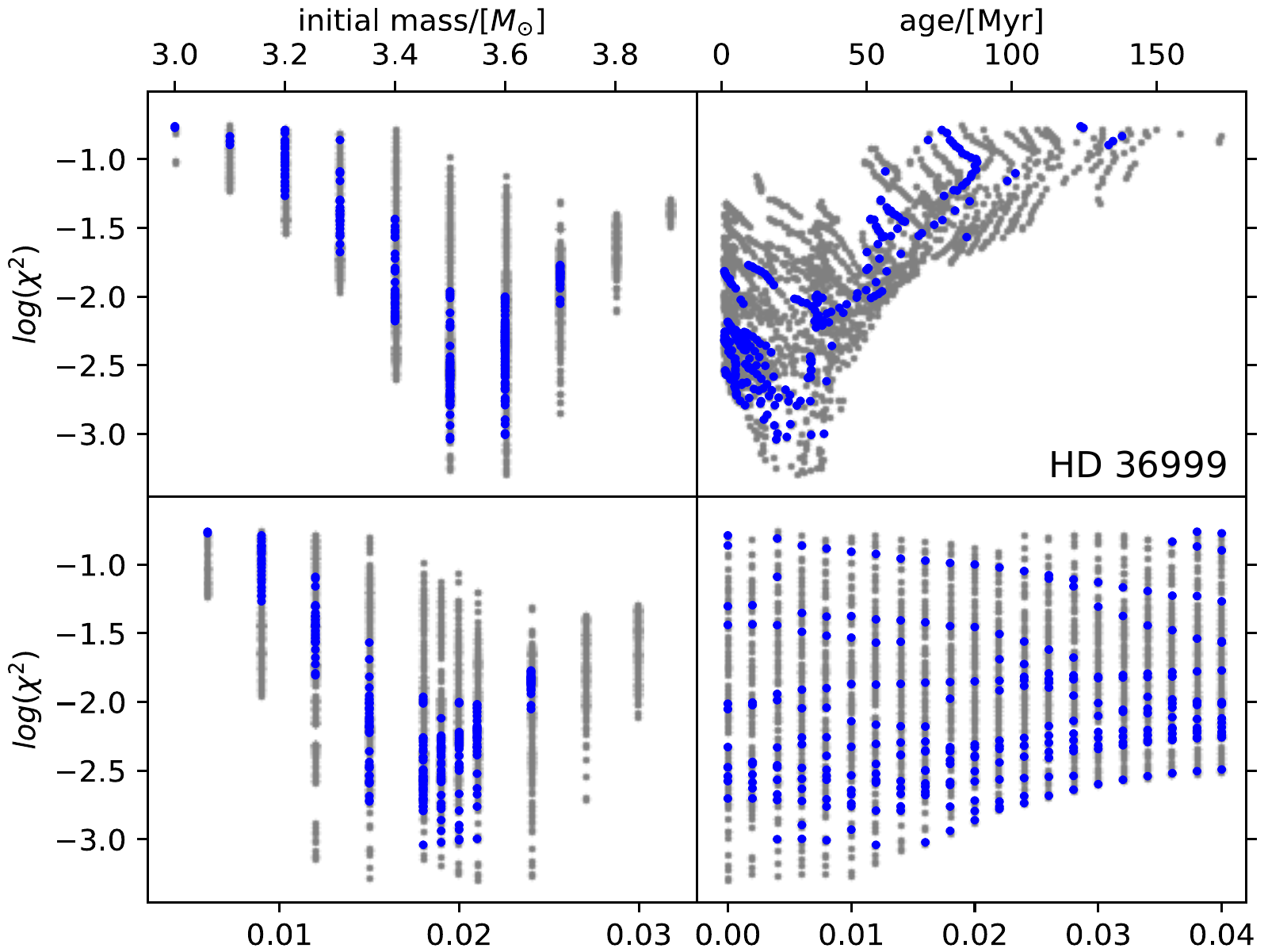}
 \includegraphics[width=0.48\linewidth]{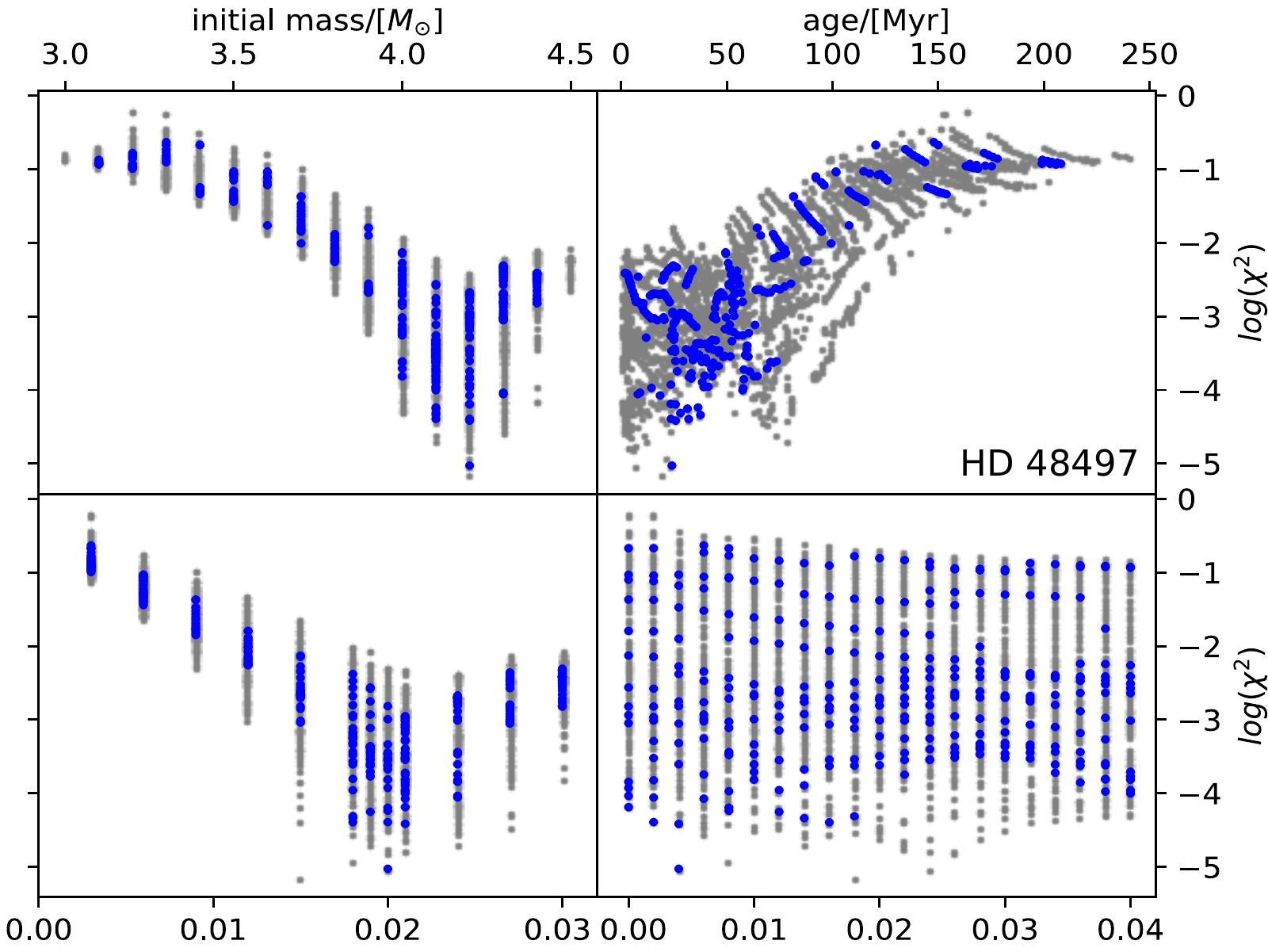}
 
 \includegraphics[width=0.48\linewidth]{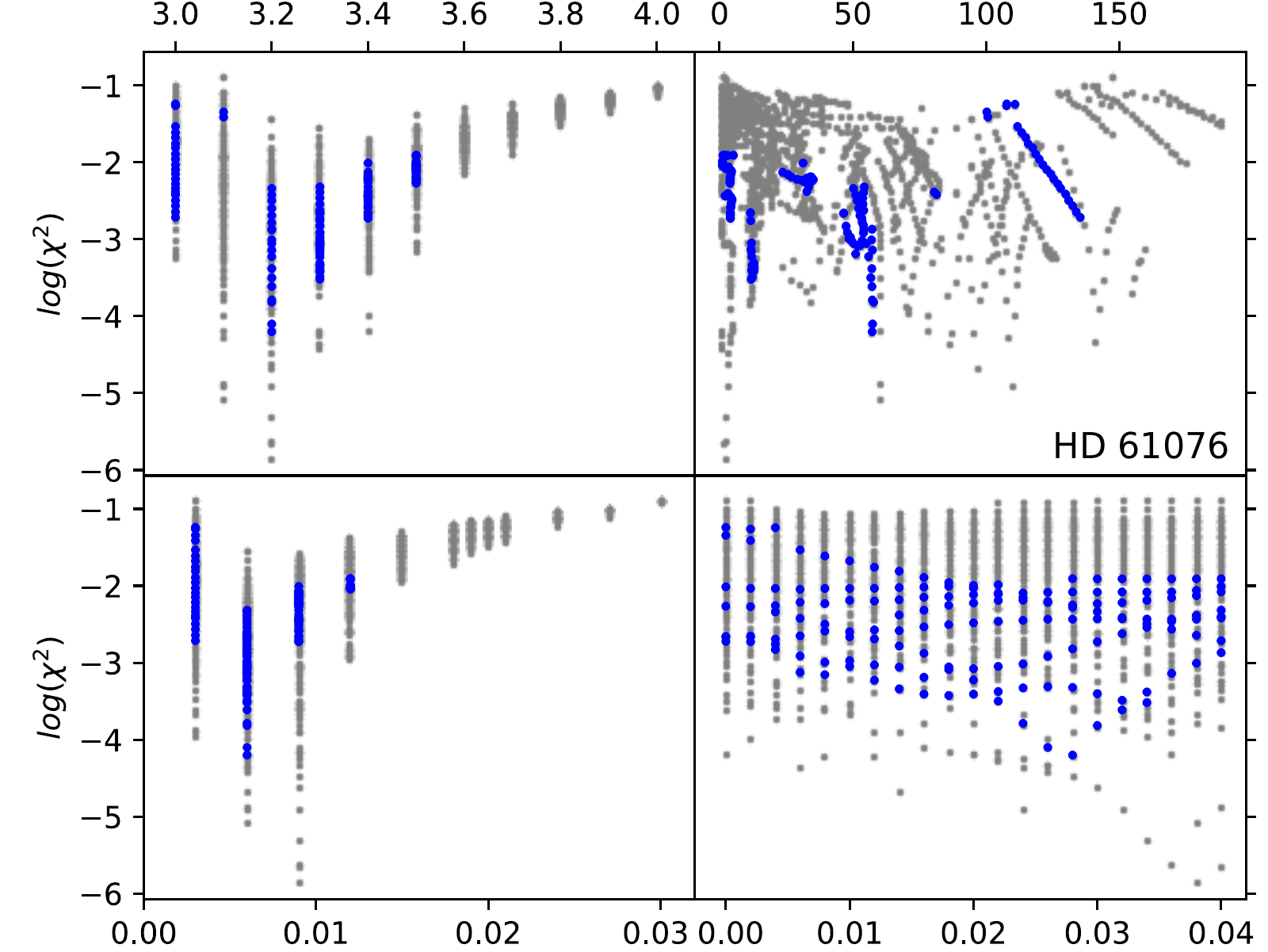}
 \includegraphics[width=0.48\linewidth]{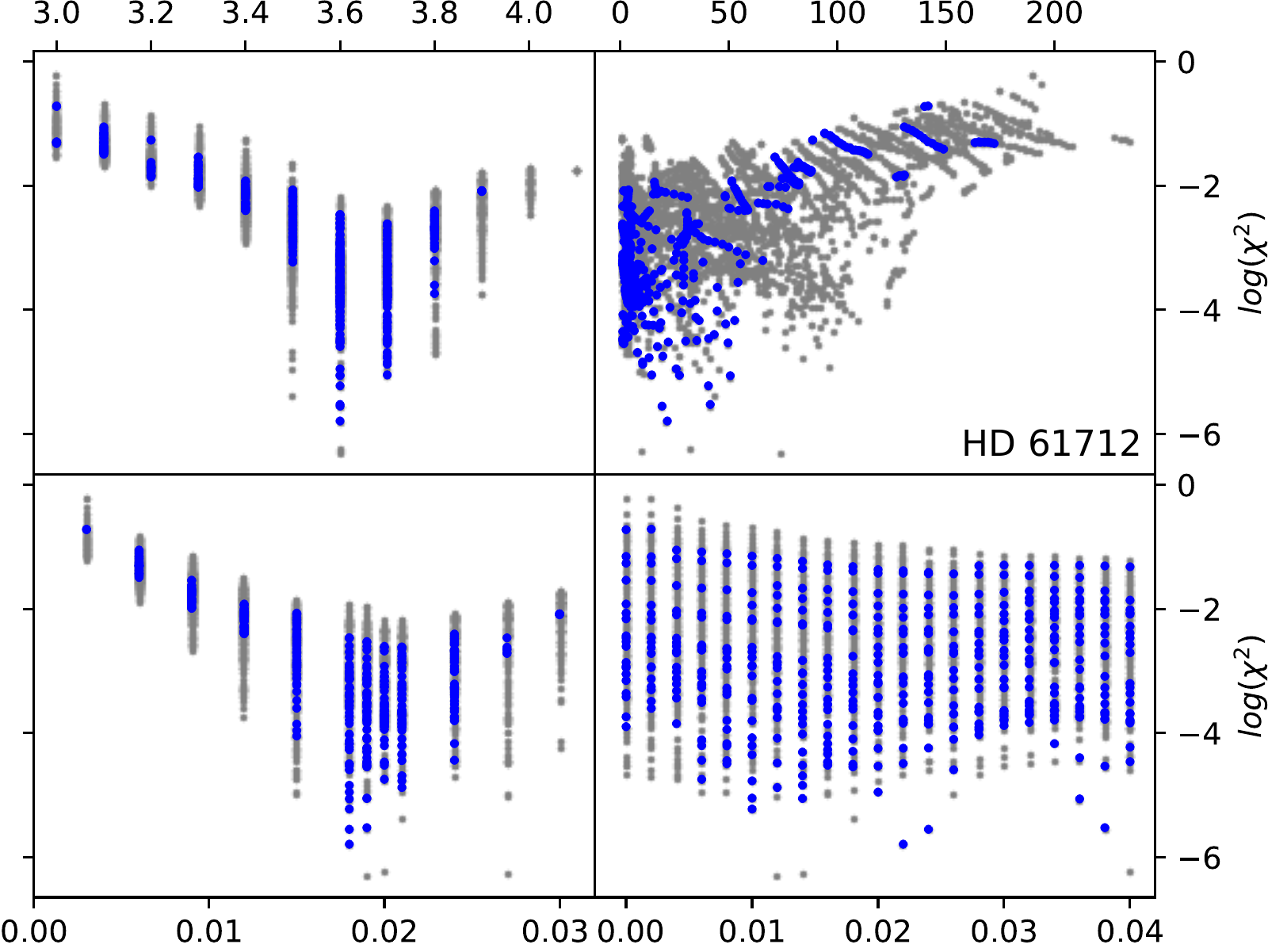}
 
 \includegraphics[width=0.48\linewidth]{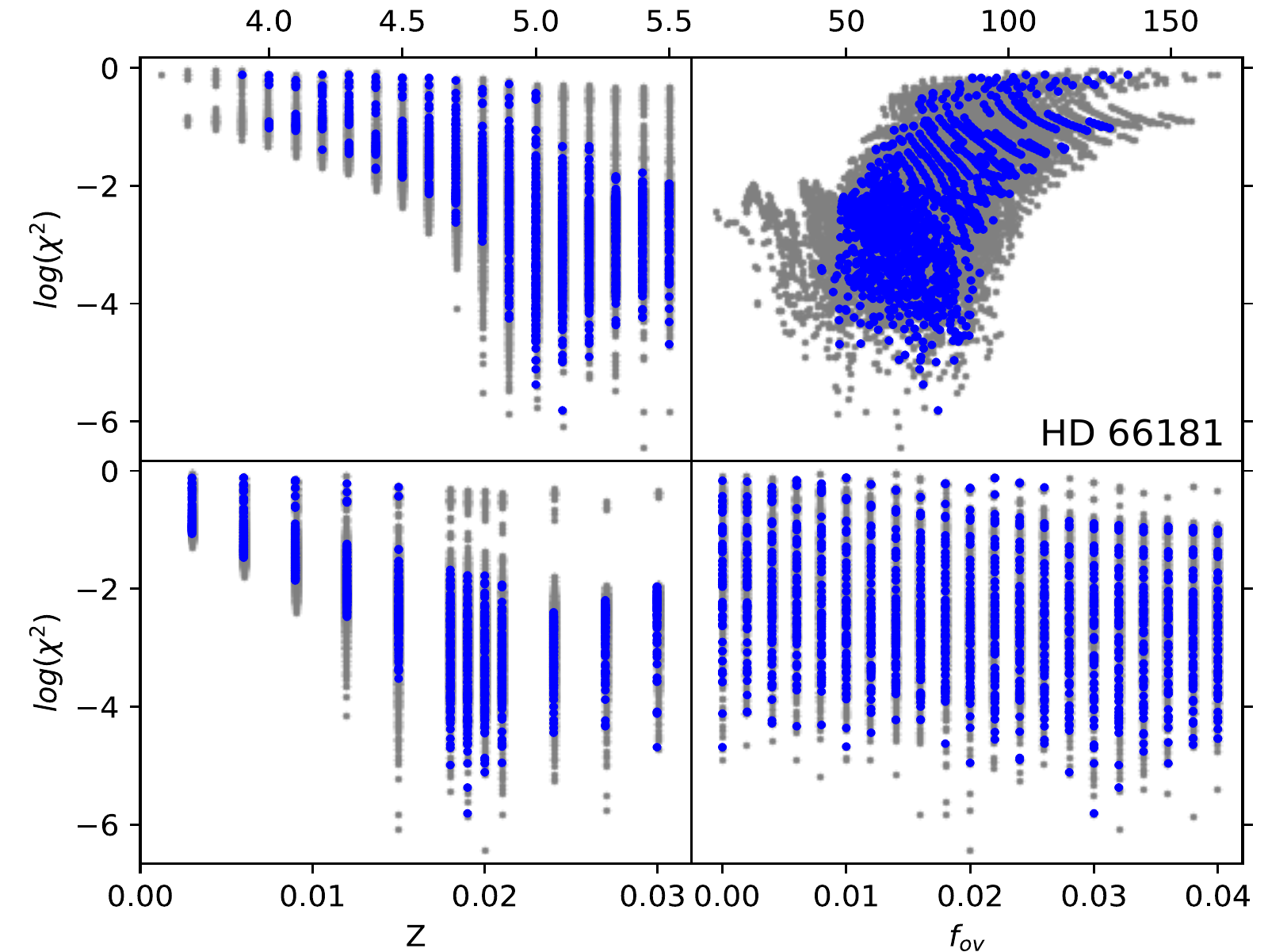}
 \includegraphics[width=0.48\linewidth]{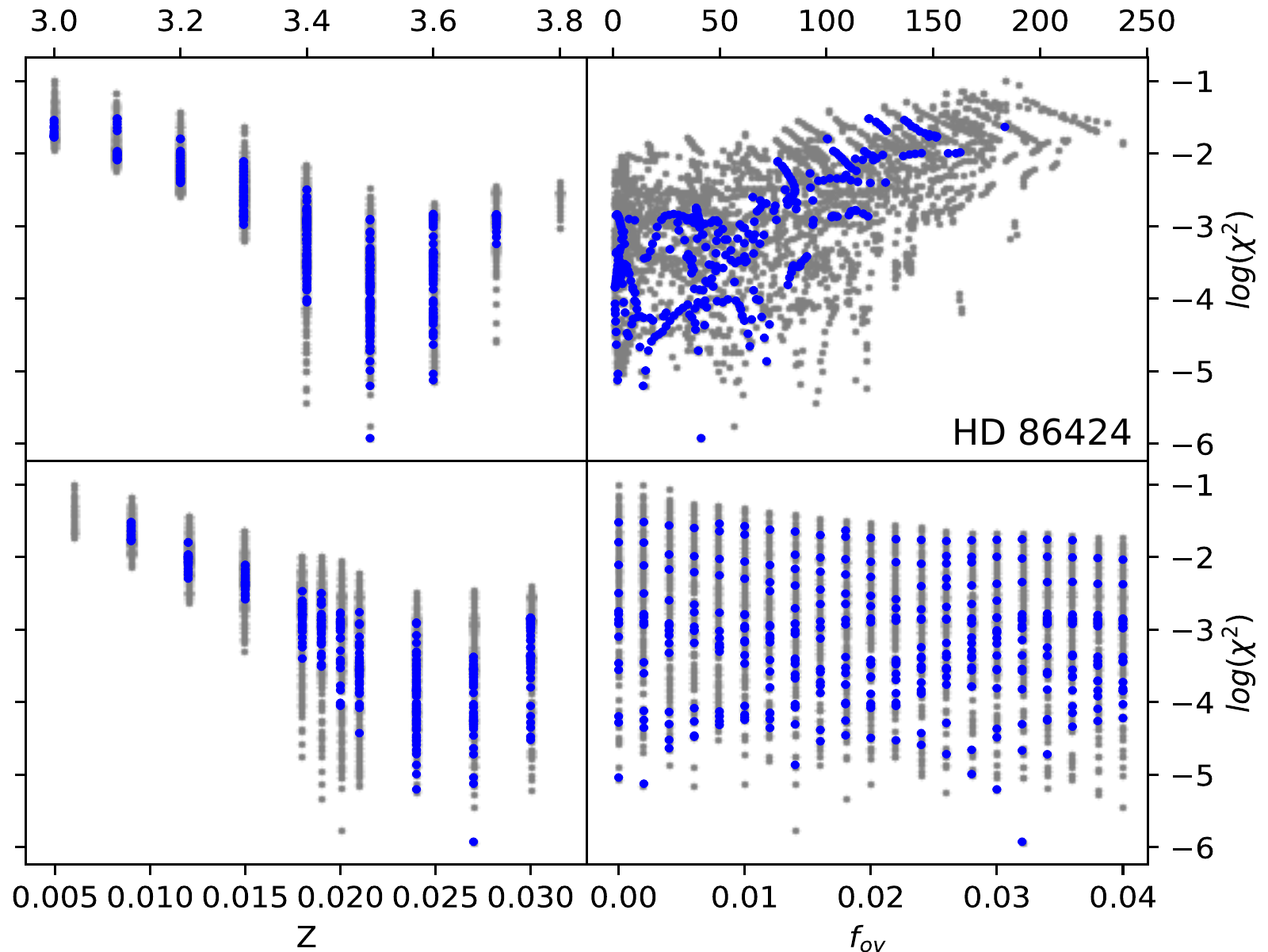}
 \caption{$\chi^2$ values of stellar models within 1$\sigma$ (blue points) and 3$\sigma$ (gray points) of the given target. Represented $\chi^2$ values were calculated for the best-fit g-mode combination as listed in Table \ref{tab:modes}. Each figure shows the $\chi^2$ distribution across all free parameters of the stellar model grid, i.e., initial mass (top left), age (top right), metallicity (bottom left), and overshooting (bottom right).}
         \label{fig:best_fit_mode_combination}
 \end{figure*}
 \begin{figure*}[h!]
 
 \includegraphics[width=0.48\linewidth]{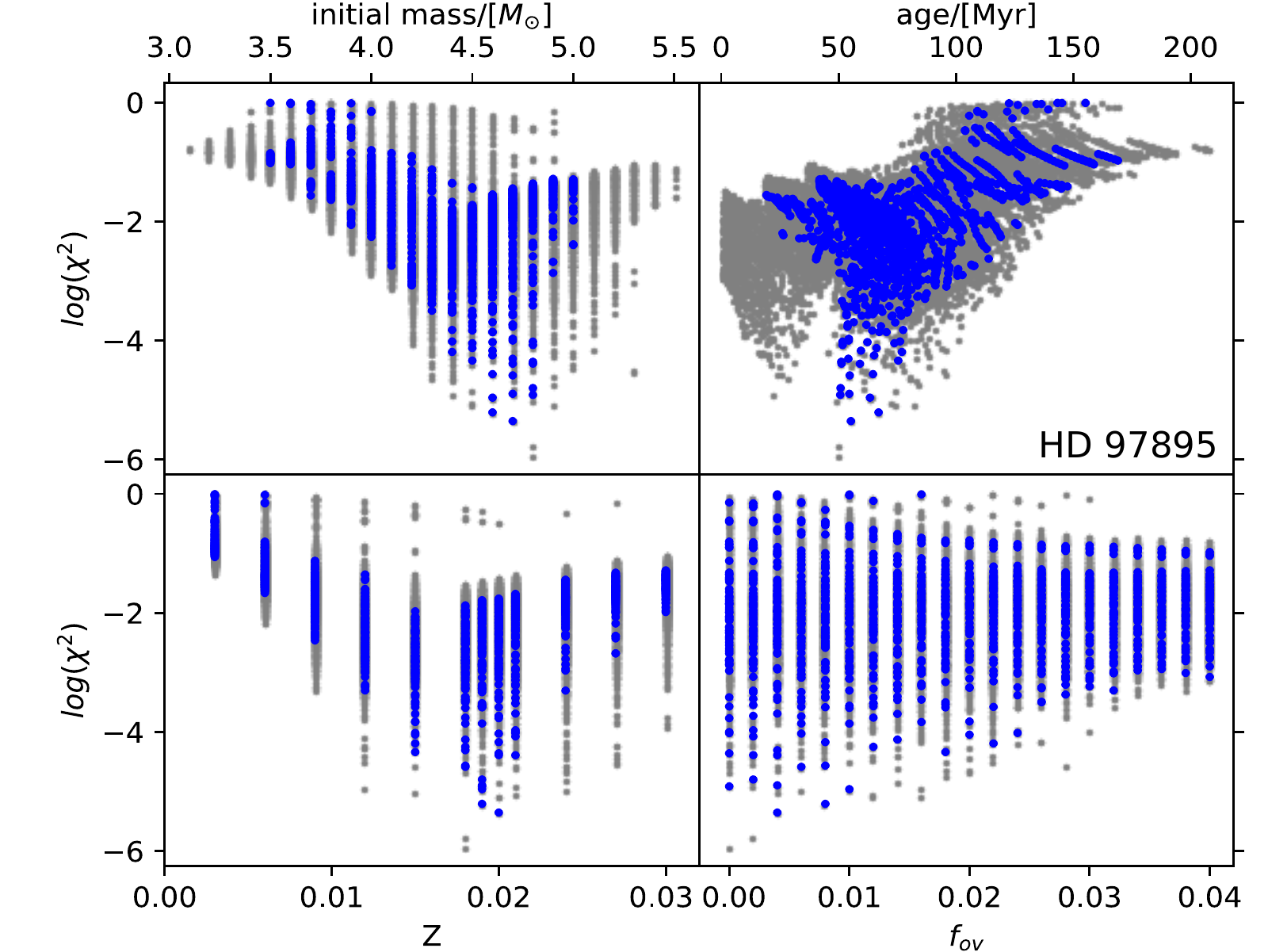}
 \includegraphics[width=0.48\linewidth]{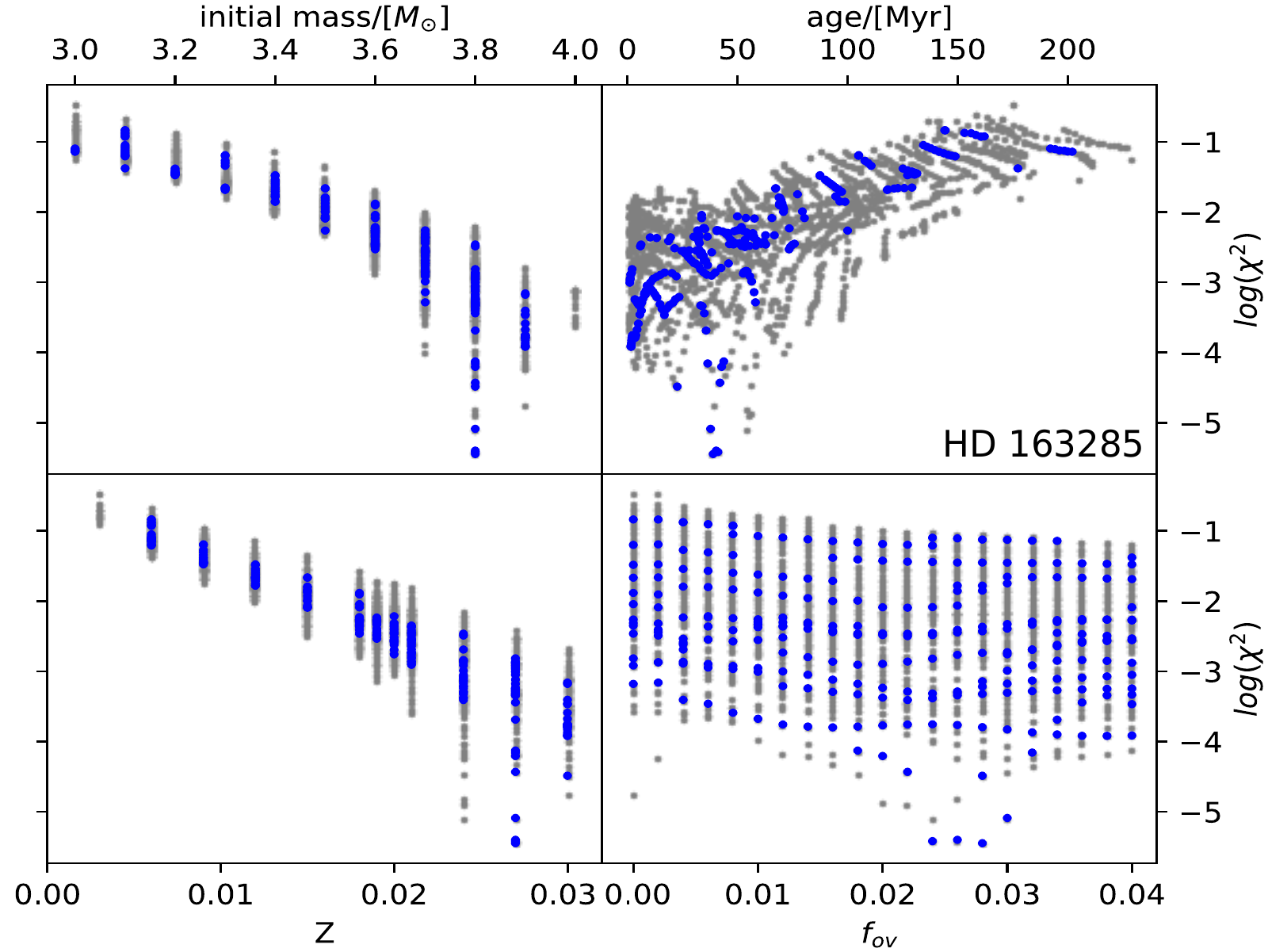}
 
 \caption{Continuation of Fig. \ref{fig:best_fit_mode_combination}}
         \label{fig:best_fit_mode_combination2}
\end{figure*}

\end{appendix}

\end{document}